\documentclass[journal,article,accept,moreauthors,pdftex,12pt,a4paper]{mdpi} 
\lastpage{x}
\doinum{10.3390/------}
\pubvolume{xx}
\pubyear{2014}
\history{Received: xx / Accepted: xx / Published: xx}

\usepackage[latin1]{inputenc}
\usepackage{amsmath}
\usepackage{array}
\usepackage{amsfonts}
\usepackage{amssymb}
\usepackage{bm}
\usepackage{bbm}
\usepackage{color}
\usepackage{graphicx}
\usepackage{algorithm2e}
\usepackage{hyperref}
\usepackage{url}
\usepackage{subfigure}

\newcommand\seq[3]{{#1}_{#2}^{#3}}
\newcommand{\bd}{\begin{document}}
\newcommand{\ed}{\end{document}}
\newcommand{\beq}{\begin{equation}}
\newcommand{\eeq}{\end{equation}}
\newcommand{\bef}{\begin{figure}}
\newcommand{\enf}{\end{figure}}
\newcommand{\bea}{\begin{eqnarray}}
\newcommand{\eea}{\end{eqnarray}}
\newcommand{\baR}{\begin{array}}
\newcommand{\eaR}{\end{array}}
\newcommand{\bc}{\begin{center}}
\newcommand{\ec}{\end{center}}
\newcommand{\ben}{\begin{enumerate}}
\newcommand{\een}{\end{enumerate}}
\newcommand{\bit}{\begin{itemize}}
\newcommand{\eit}{\end{itemize}}
\newcommand{\su}{\section}
\newcommand{\ssu}{\subsection}
\newcommand{\sssu}{\subsubsection}
\newcommand\ssssu[1]{\textbf{#1}}
\newcommand{\nid}{\noindent}
\newcommand{\nnb}{\nonumber}
\newcommand\bloc[2]{{\omega}_{#1}^{#2}}
\newcommand\blocp[2]{{\omega'}_{#1}^{#2}}
\newcommand{\un}{\mathbbm{1}}
\newcommand{\setZ}{\mathbbm{Z}}
\newcommand{\Exp}[1]{\mathbb{E}\bra{#1}}
\newcommand{\setN}{\mathbbm{N}}
\newcommand{\setR}{\mathbbm{R}}
\newcommand{\pare}[1]{\left(\, #1 \, \right)}
\newcommand{\scal}[2]{\langle\, #1 | #2 \, \rangle}
\newcommand{\bra}[1]{\left[\, #1 \, \right]}
\newcommand{\brac}[2]{\left[\, #1 \, | \, #2 \right]}
\newcommand{\Set}[1]{\left\{\, #1 \, \right\}}
\newcommand{\Setc}[2]{\left\{\, #1 \, ; \, #2 \right\}}
\newcommand{\Prob}[1]{P\left[\, #1 \, \right]}
\newcommand{\pTo}[1]{\pi^{(T)}_\omega\bra{\, #1 \, }}
\newcommand{\pTos}[1]{\pi^{(T)}_{\omega_{s}}\bra{\, #1 \, }}
\newcommand{\pT}{\pi^{(T)}_\omega}
\newcommand{\Probex}[1]{P^{(T)}\left[\, #1 \, \right]}
\newcommand{\probc}[2]{P[#1 \, \left| \, #2 \right.]}
\newcommand{\Probc}[2]{P\bra{#1 \, \left| \, #2 \right.}}
\newcommand{\Probcex}[2]{P^{(T)}\bra{#1 \, \left| \, #2 \right.}}
\newcommand{\Probcp}[2]{P^{(ap)}\bra{#1 \, \left| \, #2 \right.}}
\newcommand{\abs}[1]{\left|\, #1 \, \right|}
\newcommand{\deq}{\stackrel {\rm def}{=}}
\newcommand{\sqsubsetneq}{\stackrel {\tiny\sqsubset}{\tiny \neq}}
\newcommand\moy[1]{\mu\bra{#1}}
\newcommand\noy[1]{\nu\bra{#1}}
\newcommand\cB{{\mathcal{B}}}
\newcommand\cC{{\mathcal{C}}}
\newcommand\cD{{\mathcal{D}}}
\newcommand\cT{{\mathcal{T}}}
\renewcommand\H{{\cH_{\blambda}}}
\newcommand{\h}{\mathtt{h}}
\newcommand{\J}{\mathtt{J}}
\newcommand\s[1]{{\cS\bra{#1}}}
\newcommand\tH{{\tilde{H}}}
\renewcommand\th{{\tilde{h}}}
\newcommand\tFi{{\tilde{\Phi}}}
\newcommand\tfi{{\tilde{\phi}}}
\newcommand\tF{{\tilde{F}}}
\newcommand\tG{{\tilde{G}}}
\newcommand\tu{{\tilde{u}}}
\newcommand\tv{{\tilde{v}}}
\renewcommand{\sb}{s_\bbeta}
\newcommand\cE{{\mathcal{E}}}
\newcommand\cG{{\cal G}}
\newcommand\cH{{\mathcal{H}}}
\newcommand\cI{{\mathcal{I}}}
\newcommand\cK{{\cal K}}
\newcommand\cL{{\cal L}}
\newcommand\cM{{\mathcal{M}}}
\newcommand\cS{{\cal S}}
\newcommand\cP{{\mathcal{P}}}
\newcommand\cO{{\cal O}}
\newcommand\cU{{\cal U}}
\newcommand\cW{{\cal W}}
\newcommand\f{{\bm{f}}}
\newcommand\be{{\bm{e}}}
\newcommand\bh{{\bm{h}}}
\newcommand\bxi{{\bm{\xi}}}
\newcommand\bn{{\bm{n}}}
\newcommand\bv{{\bm{v}}}
\newcommand\bF{{\bm{F}}}
\newcommand\bo{{\bm{o}}}
\newcommand\blambda{{\bm{\lambda}}}
\newcommand\bbeta{{\bm{\beta}}}
\newcommand\beps{{\bm{\epsilon}}}
\newcommand\Eta{{\bm{\eta}}}
\newcommand\gk[1]{g_k\pare{#1}}
\newcommand\gLk{g_{L,k}}
\newcommand\akj[1]{\alpha_{kj}(#1)}
\newcommand\sif[1]{\seq{\omega}{#1}{D}}
\newcommand\Xk[1]{X_k({#1})}
\newcommand\Ik[1]{I_k({#1})}
\newcommand\Skj[1]{S_{kj}({#1})}
\newcommand\X[2]{X_{#1}\pare{#2}}
\newcommand\et[2]{\eta_{#1}\pare{#2}}
\newcommand\tO[2]{\tau_{#1}\pare{#2}}
\newcommand\Vd[2]{\mathcal{V}^{(det)}_{#1}\pare{#2}}
\newcommand\XkD{\Xk{\bloc{0}{D-1}}}
\newcommand\Xkr{X_k^r}
\newcommand\tLk{\tau_{L,k}}
\newcommand\moyc[2]{\mu\bra{#1 \, | \, #2}}
\newcommand\Vkdet[1]{V_k^{(det)}({#1})}
\newcommand\Vksyn[1]{V_k^{(syn)}({#1})}
\newcommand\Vkext[1]{V_k^{(ext)}({#1})}
\newcommand\Vknoise[1]{V_k^{(noise)}({#1})}
\newcommand\sk[1]{\sigma_k({#1})}
\newcommand\skr{\sigma_k^r\pare{\bloc{0}{D-1}}}
\newcommand{\ent}[1]{\left[\, #1 \, \right]}
\newcommand\vm[1]{var_m\left[#1 \right]}
\newcommand\eg[1]{\stackrel {\rm #1}{=}}
\newcommand\tk[1]{\tau_k\pare{#1}}
\newcommand\tko{\tau_k(t,\omega)}
\newcommand\cBm[1]{\cB^{-1}\pare{#1}}
\newtheorem{theorem}{Theorem}
\newcommand{\bth}{\begin{theorem}}
\newcommand{\enth}{\end{theorem}}
\newcommand\omD[1]{\seq{\bra{\omega^{(#1)}}}{0}{D-1}}
\newcommand\omz[1]{\omega^{(#1)}(D)}
\newcommand\om[1]{\omega^{(#1)}}
\newcommand\Om[1]{\Omega^{(#1)}}
\newcommand\skjq{s_{kj;q}}
\newcommand\ikq{i_{k;q}}
\newcommand\skjqs{p_{kj;q_s}}
\newcommand\ikqs{i_{k;q_s}}
\newcommand\ikqsu{i_{k;q_{s_u}}}
\newcommand\ikqu{i_{k;q_u}}
\newcommand\xkq{x_{k;q}}
\newcommand\skjqv{s_{kj;q_{v}}}
\newcommand\tjr{t_j^{(r)}(\omega)}
\renewcommand\S[2]{S_{#1}\pare{#2}}
\newcommand{\im}{Im}
\newcommand\rpf{R}
\newcommand\rpfc[1]{R\pare{#1}}
\newcommand\lpf{L}
\newcommand\lpfc[1]{L\pare{#1}}
\newcommand{\zb}{\zeta_{\bbeta}}
\newcommand{\etab}{\eta_{\bbeta}}
\newcommand{\mex}{\mu^{(ex)}}
\newcommand{\phex}{\phi^{(ex)}}
\newcommand{\Phex}{\Phi^{(ex)}}
\newcommand{\Hex}{\cH^{(ex)}}
\newcommand{\hex}{h^{(ex)}}
\newcommand{\siex}{\sigma^{(ex)}}
\newcommand{\cn}{\cC^{\ast}}
\newcommand\p[1]{\cP\bra{#1}}
\newcommand\w[2]{w_{#1}^{(#2)}}
\newcommand\Gb{\cG_\bbeta}
\newcommand\K[2]{\cK_{\small{#1; \,  #2}}}
\renewcommand\d[2]{\delta_{\small{#1; \,  #2}}}
\newcommand\blp[3]{\omega^{#3,(#1)}_{#2}}
\newcommand\blpb[3]{\overline{\omega^{#3,(#1)}_{#2}}}
\newcommand\moyex[1]{\mu^{(ex)}\bra{#1}}
\newcommand\E[1]{{\cal E}_{#1}}
\newcommand{\tc}{\tau(\cC)}
\newcommand{\notsqsupset}{\cancel{\sqsupset}}
\newcommand{\sm}{\sigma^{-1}}
\newcommand\hP{\hat{\cP}}
\newcommand\bdelta{{\bm{\delta}}}
\newcommand\Hp{{\cH_{\blambda'}}}
\newcommand\Ha{{\cH_{\blambda^\ast}}}
\newcommand\mb{\mu_\blambda}
\newcommand\mB[1]{\mu_\blambda\bra{#1}}
\newcommand\mbap{\mu^{ap}_{\blambda}}
\newcommand\mbp{\mu_{\blambda^{\prime}}}
\newcommand\mbs{\mu_{\blambda^\ast}}
\newcommand\mk{m_k}
\newcommand\ml{m_l}
\newcommand\mj{m_j}
\newcommand\m{\textbf{m}}
\newcommand\pres[1]{\cP\bra{#1}}
\newcommand\Zb{Z_{\blambda}}

\Title{Parameters estimation for spatio-temporal maximum entropy distributions: application to neural spike trains}

\Author{Hassan Nasser $^{1,*}$ and Bruno Cessac $^{1,*}$}

\address{%
$^{1}$ INRIA, 2004 route de lucioles, 06560, Sophia-Antipolis, France
}
\corres{\\(Hassan.Nasser;Bruno.Cessac)@inria.fr}

\abstract{We propose a numerical method to learn Maximum Entropy (MaxEnt) distributions with  \textit{spatio-temporal constraints} from experimental spike trains. This is an extension of two papers [10] and [4] who proposed the estimation of parameters where only \textit{spatial constraints} were taken into account.  The extension we propose allows to properly handle memory effects in spike statistics, for large sized neural networks.}


\begin{document}

\section{Introduction}
With the evolution of Multi-Electrode Arrays (MEA) acquisition techniques, it is currently possible to simultaneously record the activity of a few hundred of neurons up to a few thousand \cite{ferrea-etal:12}. Stevenson et al \cite{stevenson-kording:2011} reported that the number of recorded neurons doubles approximately every 8 years. However, beyond the mere recording of an increasing number of neurons, there is a need to extract relevant information from data in order to understand the underlying dynamics of the studied network, how it responds to stimuli and how spike train response encodes these stimuli. In the realm of spike trains analysis this means having efficient spike sorting techniques \cite{marre-amodei-etal:12, hill-etal:11, litke-etal:04, quiroga-etal:04}, but also efficient methods to analyze  spike statistics. The second aspect requires using canonical statistical models whose parameters have to be tuned 
("learned") from data.

The Maximum Entropy method (MaxEnt) offers a way to selecting canonical statistical models from first principles. Having its root in statistical physics, MaxEnt consists of fixing a set of constraints, determined as the empirical average of features measured from the spiking activity. Maximizing the statistical entropy given those constraints provides a unique probability, called a Gibbs distribution, which approaches at best data statistics in the following sense: among all probability distributions which match the constraints this is the one which has the smallest Kullback-Leibler divergence with the data (\cite{csiszar:74}). Equivalently, it satisfies the constraints without adding additional assumption on statistics \cite{jaynes:57}. 

Most studies have focused on describing properly the statistics of \textit{spatially} synchronized patterns of neuronal activity without considering time-dependent patterns and memory effects. In this setting pairwise models \cite{schneidman-berry-etal:06,pillow-shlens-etal:08} or extensions with triplets and quadruplets interactions \cite{ganmor-segev-etal:11b}, \cite{ganmor-segev-etal:11a}, \cite{tkacik-schneidman-etal:09} were claimed to correctly fit $\approx 90$ to $99\%$ of the information. However, considering now the capacity of these models to correctly reproduce \textit{spatio-temporal} spike patterns, the performances drop-off dramatically, especially in the cortex \cite{tang-jackson-etal:08, marre-boustani-etal:09} or in the retina \cite{vasquez-marre-etal:12}.

Taking into account spatio-temporal patterns requires to introduce memory in statistics, described as a Markov process. MaxEnt extends easily to this case (see section \ref{SMaxEnt} and references therein for a short description) producing Gibbs distributions in the spatio-temporal domain. Moreover, rigorous mathematical methods are available to fit the parameters of the Gibbs distribution \cite{vasquez-marre-etal:12}. However, the main drawback of these methods is the huge computer memory they require, preventing their applications to large scale neural networks. Considering a model with memory depth $D$ (namely, the probability of a spike pattern at time $t$ depends on the spike activity in the interval $[t-D, t-1]$), there are $2^{N(D+1)}$ possible patterns. The method developed in \cite{vasquez-marre-etal:12} requires to handle a matrix of size $2^{N(D+1)} \times 2^{N(D+1)}$. So, it becomes intractable for $N(D+1) > 20$.

In this paper, we propose an alternative method to fit the parameters of a spatio-temporal Gibbs distribution with larger values of the product $N(D+1)$. We have been able to go up to $N(D+1)$ ($\sim 120$) on a small cluster (64 processors AMD Opteron(tm)  $2300$ MHz). The method is based on \cite{dudik-phillips-etal:04} and \cite{broderick-dudik-etal:07} who proposed the estimation of parameters in \textit{spatial} Gibbs distributions. The extension in the \textit{spatio-temporal} domain is not straightforward, as we show, but it carries over to the price of some modifications.  Combined with parallel Montecarlo computing developed in \cite{nasser-marre-etal:13} this provides a numerical method allowing to handle Markovian spike statistics with spatio-temporal constraints. 

The paper is organized as follow. In section \ref{background}, we recall the theoretical background for spike train with Gibbs distribution. We discuss both spatial and spatio-temporal case. In the next section, \ref{inferring}, we explain the method to fit the parameters of MaxEnt distributions. As we mathematically show, the convex criterion used by \cite{dudik-phillips-etal:04} still applies for spatio-temporal constraints. However, the method used by \cite{broderick-dudik-etal:07} to avoid recomputing the Gibbs distribution at each parameters change cannot be directly used and has to be adapted using a Linear Response scheme.
In the last section, \ref{tests}, we show benchmarks  evaluating the performance of this method and  discuss the computational obstacles that we encountered. We made tests  with both synthetic and real data. Synthetic data were generated from known probability distributions using a Montecarlo method. Real data corresponds to spike trains obtained from retinal ganglion cells activity (courtesy of M.J. Berry and O. Marre). The method shows a satisfying performance in the case of synthetic data. Real data analysis is not systematic but instead used as an illustration and comparison with the paper of Schneidman et al. 2006 (\cite{schneidman-berry-etal:06}). As we could see in the example, the performance on real data, although satisfying, is affected by the large number of parameters in the distribution, consequence of the choice to work with canonical models (Ising, pairwise with memory). This effect is presumably not related to our method but to a standard problem in statistics. 

Some of our notations might be not usual to some readers. Therefore, we added a list of symbols at the end of the paper.

\section{Gibbs distributions in the spatio-temporal domain}
\label{background}
\ssu{Spike trains and observables} \label{SSpikes}

\sssu{Spike trains} 
We consider the joint activity of $N$ neurons, characterized by the emission of action potentials ("spikes"). We  assume that there is a minimal time scale, $\delta$, set to $1$ without loss of generality such that a neuron can at most fire a spike within a time window of size $\delta$. This provides a time discretization labeled with an integer time $n$. Each neuron activity is then characterized  by a binary variable\footnote{We use the notation $\omega$ to differentiate our binary variables $\in \Set{0,1}$ to the notation $\sigma$ or $S$ used for ``spins'' variables $\in \Set{-1,1}$.} $\omega_k(n)=1$ if  neuron $k$ fires at time $n$ and $\omega_k(n)=0$ otherwise. 

The state of the entire network in time bin $n$ is thus described by a vector  $\omega(n) \deq \bra{\omega_k(n)}_{k=1}^{N}$, called a \textit{spiking pattern}.
A {\em spike block} is  a consecutive sequence of spike patterns $\omega_{n_1}^{n_2}$, representing the activity of the whole network between two instants $n_1$ and $n_2$.
$$\bloc{n_1}{n_2} = \Set{\omega(n)}_{\{n_1 \leq n \leq n_2\}}.$$
 The \textit{time-range} (or "range") of a block $\bloc{n_1}{n_2}$ is $n_2-n_1+1$, the number of time steps from $n_1$ to $n_2$. 
Here is an example of a spike block with $N=4$ neurons and range $R=3$:
$$
\tiny{\bra{
\begin{array}{ccccccc}
0 & 1 & 1\\
0 & 0 & 1\\
1 & 0 & 1\\
1 & 1 & 1
\end{array}
}}
$$ 
A \textit{spike train} or \textit{raster} is a spike block $\bloc{0}{T}$ from some initial time $0$
to some final time $T$.
To alleviate notations we simply write $\omega$ for a spike train. We note $\Omega$ 
the set of spike trains.

\sssu{Observables} \label{Sobs}

An \textit{observable} is a function $\cO$ which associates a real number $\cO(\omega)$ to a spike train. In the realm of statistical physics common examples of observables are the energy or the number of particles (where $\omega$ would correspond e.g. to a spin configuration). In the context of neural networks examples are the number of neuron firing at a given time $n$, $\sum_{k=1}^N \omega_k(n)$, or the function $\omega_{k_1}(n_1) \omega_{k_2}(n_2)$ which is $1$ if neuron $k_1$ fires at time $n_1$ and neuron $k_2$ fires at time $n_2$ and is $0$ otherwise.

Typically, an observable does not depend on the full raster, but only on a sub-block of it. The \textit{time-range} (or "range") of an observable is the minimal integer $R>0$ such that, for any raster $\omega$, $\cO(\omega) = \cO\pare{\bloc{0}{R-1}}$. The range of the observable $\sum_{k=1}^N \omega_k(n)$ is $1$; the range of $\omega_{k_1}(n_1) \omega_{k_2}(n_2)$ is $n_2 - n_1 +1$.
From now on, we restrict to observables of range $R$, fixed and finite. We set $D=R-1$.

An observable is \textit{time-translation invariant} if, for any time $n>0$ we have $\cO\pare{\bloc{n}{n+D}} \equiv \cO\pare{\bloc{0}{D}}$ whenever $\bloc{n}{n+D}=\bloc{0}{D}$. The two examples above are time-translation invariant. The observable $\lambda(n_1) \omega_{k_1}(n_1) \omega_{k_2}(n_2)$, where $\lambda$ is a real function of time, is not time-translation invariant. Basically, time-translation invariance means that $\cO$ does not depend explicitly on time. We focus on such observables from now on.

\sssu{Monomials} \label{SMonomials}

Prominent examples of time-translation invariant observables with range $R$ are products of the form:
\beq\label{DefMon}
m_{p_1, \dots, p_r}(\omega)\deq \prod_{u=1}^r \omega_{k_u}(n_u).
\eeq
where $p_u, \, u=1 \dots r$ are pairs of spike-time events $(k_u, n_u)$, $k_u=1 \dots N$ being the neuron index, and $n_u=0 \dots D$ being the time index. Such an observable,
called \textit{monomial}, takes therefore values in $\Set{0,1}$ and is $1$ if and only if $\omega_{k_u}(n_u)=1$,
$u=1 \dots r$ (neuron $k_1$ fires at time $n_1$, $\dots$, neuron $k_r$ fires at time $n_r$).  
A monomial is therefore a binary observable that represents the logic-AND operator applied to a prescribed set of  neuron spikes events.

We allow the extension of the definition (\ref{DefMon}) to the case where the set of pairs ${p_1, \dots, p_r}$ is empty and we set $m_\emptyset=1$. 
For a number $N$ of neurons and a time range $R$ there are
thus $2^{N \,R}$ such possible products. 
Any observable of range $R$ can be represented as a linear combination of products (\ref{DefMon}). Monomials constitute therefore a canonical basis for observable representation. To alleviate notations, instead of labeling monomials by a list of pairs, as in (\ref{DefMon}), we shall label them by an integer index $l$.

\sssu{Potential} \label{SPotential}

Another prominent example of observable is the function called "energy" or \textit{potential} in the realm of the MaxEnt. Any potential of range $R$ can be written as a linear combination of the $2^{NR}$ possible monomials (\ref{DefMon}):
\beq\label{Hh}
\H \, =\sum_{l=1}^{2^{NR}} \lambda_l m_l,
\eeq
where some coefficients $\lambda_l$ in the expansion may be zero. Therefore, by analogy with spin systems, monomials somewhat constitute spatio-temporal interactions between neurons: the monomial $\prod_{u=1}^r \omega_{k_u}(n_u)$ 
contributes to the total energy $\H(\omega)$ of the raster $\omega$ if and only if  neuron $k_1$ fires at time $n_1$, $\dots$, neuron $k_r$ fires at time $n_r$ in the raster $\omega$.  The number of pairs in
a monomial (\ref{DefMon}) defines the degree of an interaction: degree $1$ corresponds to "self-interactions", degree $2$ to pairwise, and so on. Typical examples of such potentials
are the Ising model \cite{schneidman-berry-etal:06,pillow-shlens-etal:08,schaub-schultz:10}:
\beq \label{HIsing}
\cH_{Ising}\pare{\omega(0)} = \sum_i \lambda_i \omega_i(0) + \sum_{ij} \lambda_{ij}\omega_i(0)\omega_j(0),
\eeq
where considered events are individual spikes and pairs of simultaneous spikes.
Another example is the Ganmor-Schneidman-Segev (GSS) model \cite{ganmor-segev-etal:11a}, \cite{ganmor-segev-etal:11b} 
\beq \label{HGSS}
\cH_{GSS}\pare{\omega(0)} = \sum_i \lambda_i \omega_i(0) + \sum_{ij} \lambda_{ij} \omega_i(0)\omega_j(0) + \sum_{ijk} \lambda_{ijk} \omega_i(0)\omega_j(0)\omega_k(0),
\eeq
where additionally to \ref{HIsing}, simultaneous triplets of spikes are considered (We restrict the form (\ref{HGSS}) to triplet although Ganmor et al were also considering quadruplets). In these two examples the potential is a function of the spike pattern at a given time. Here, we choose this time equal to $0$, without loss of generality, since we are considering time-translation invariant potentials. More generally, the form (\ref{Hh}) affords
the consideration of spatio-temporal neurons interactions: this allows us to introduce delays, memory and causality in spike statistics estimation. A simple example is a pairwise model with delays such as:
\beq\label{HPR}
\cH_{PR}\pare{\bloc{0}{D}} = \sum_i \lambda_i \omega_i(D) + \sum_{s=0}^{D}\sum_{ij} \lambda_{ij}^{s}\omega_i(0)\omega_j(s),
\eeq
where 'PR' stands for 'Pairwise with range R', takes into account the events where neuron $i$ fires $s$ time steps after a neuron $j$ with $s=0 \dots D$.

\ssu{The Maximum Entropy Principle}\label{SMaxEnt}

Assigning equal probabilities (uniform probability distribution) to possible outcomes  goes back to Laplace and Bernoulli (\cite{garibaldi-penco:85}) ("principle of insufficient reason"). Maximizing the statistical entropy {\it without} constraints is equivalent to this principle. In general, however, one has some knowledge about data, typically characterized by empirical average of prescribed observables (e.g. for spike trains, firing rates, probability that a fixed group of neurons fire at the same time, probability that $K$ neurons fire at the same time \cite{tkacik-marre-etal:13}): this constitutes a set of constraints.
The Maximum Entropy Principle (MaxEnt) is a method to obtain, from the observation of a statistical sample, a probability distribution that approaches at best the statistics of the sample, taking into account these constraints without additional assumptions  \cite{jaynes:57}.
 Maximizing the statistical entropy given those  constraints provides a distribution as far as possible from the uniform and as close as possible to the empirical distribution.  For instance,  considering the empirical mean and variance of the sample of a random variable as constraints results in a Gaussian distribution.

Although some attempts have been made to extend MaxEnt to non stationary data \cite{jaynes:78,jaynes:80,jaynes:85,otten-stock:10} it is mostly applied in the context of stationary statistics: the average of an observable does not depend explicitly on time. We shall work with this hypothesis. In its simplest form, the MaxEnt also assumes that the sample has no memory: the probability of an outcome at time $t$ does not depend on the past. We first discuss the MaxEnt in this context in the next section, before considering the case of processes with memory in the section \ref{SMaxEntGen}.

\sssu{Spatial constraints}\label{SMaxEntSimple}

In our case, the natural constraints are represented by the empirical probability of occurrence of characteristic spike events in the spike train, or, equivalently, by the average of specific monomials. Classical examples of constraints are the probability that a neuron fires at a given time (firing rate) or the probability that two neurons fire at the same time. 
For a raster $\omega$ of length $T$ we note $\pT$ the empirical distribution, and $\pTo{\cO}$ the empirical average of the observable $\cO$ in the raster $\omega$. For example, the empirical firing rate of neuron $i$ is 
$\pTo{\omega_i} = \frac{1}{T} \sum_{n=0}^{T-1} \omega_i(n)$, the empirical probability that two neurons $i,j$ fire at the same time is $\pTo{\omega_i \omega_j} = \frac{1}{T} \sum_{n=0}^{T-1} \omega_i(n)\omega_j(n)$ and so on. Given a set of $L$ monomials $m_l$,
their empirical average, $\pTo{m_l}$, measured in the raster $\omega$, constitute a set of constraints shaping the sought probability distribution.
We consider here monomials corresponding to events occurring \textit{at the same time},
i.e. $m_l(\omega) \equiv m_l\pare{\omega(0)}$ postponing to section \ref{SMaxEntGen} the general case of events occurring at distinct times.

In this context, the MaxEnt problems is stated as follows. Find a \textit{probability distribution} $\mu$ that maximizes the entropy:
\beq\label{Ssimple}
\s{\mu}\,=\, - \, \sum_{\omega(0)} \moy{\omega(0)} \log \moy{\omega(0)},
\eeq
(where the sum holds on the $2^N$ possible spike patterns $\omega(0)$),
 given the constraints:
\begin{equation}\label{Constraints}
 \moy{m_l} = \pTo{m_l}, \, l=1 \dots L.
\end{equation}
The average of monomials, predicted by the statistical model $\mu$ (noted here $\moy{m_l}$), must be 
equal to the average $\pTo{m_l}$ measured in the sample. There is, additionally, the probability normalization constraint:
\beq\label{Normalisation}
\sum_{\omega(0)} \moy{\omega(0)}=1
\eeq

This provides a variational problem
\begin{equation}\label{VarPrincWeak}
\mu = \arg \max_{\nu \in \cM} \bra{\s{\nu} + \lambda_0 \pare{\sum_{\omega(0)} \noy{\omega(0)}-1} + \sum_{l=1}^L \lambda_l \pare{\noy{m_l}-\pTo{m_l}}}                                                                                                                                                                                                                                                                                                                                                                                                                                                                                                                                                                                                                                                                                                                                   
\end{equation}
where $\cM$ is the set of (stationary) probabilities on spike trains. One searches, among all stationary probabilities $\nu \in \cM$, the one which maximizes the rhs of (\ref{VarPrincWeak}). There is a unique such probability, $\mu = \mu_{\blambda}$, provided $N$ is finite and $\lambda_l > -\infty$. This probability depends on the parameters $\blambda$.

Stated in this form the MaxEnt is a Lagrange multipliers problem. The sought probability distribution is the classical Gibbs distribution:
\beq\label{Gibbs_weak}
\mB{\omega(0)} = \frac{1}{\Zb}e^{\H\bra{\omega(0)}},
\eeq
where $\Zb=\sum_{\omega(0)} e^{\H\bra{\omega(0)}}$ is the \textit{partition function}, whereas $\H\bra{\omega(0)} \, =\sum_{l=1}^{L} \lambda_l m_l\bra{\omega(0)}$. Note that the time index (here $0$) does not play a role since we have assumed $\mb$ to be stationary (time-translation invariant).

The value of $\lambda_l$s is fixed by the relation:
\beq\label{moy_weak}
 \mb{(m_l)} = \, \frac{\partial \, \log \Zb}{\partial \lambda_l} = \pTo{m_l}, \, l=1 \dots L.
\eeq
Additionally, note that the matrix $\frac{\partial^2 \, \log \Zb}{\partial \lambda_l \, \partial \lambda_{l'}}$ is positive. This ensures the convexity of the problem and the uniqueness of the solution of the variational problem.

Note that we do not expect  in general $\mb$ to be equal to the (hidden) probability shaping the observed sample. It is only the closest one satisfying the constraints (\ref{Constraints}) \cite{csiszar:74}.
The notion of closeness is related to the Kullback-Leibler divergence, defined in the next section.\\

It is easy to check
that the Gibbs distribution (\ref{Gibbs_weak}) obeys:
\beq\label{mu_ind}
\mB{\bloc{n_1}{n_2}} = \prod_{n=n_1}^{n_2} \mB{\omega(n)},
\eeq
for any spike block $\bloc{n_1}{n_2}$. Indeed, the potential of the spike
block $\bloc{n_1}{n_2}$ is $\H\pare{\bloc{n_1}{n_2}}=\sum_{n=n_1}^{n_2} \H\pare{\omega(n)}$
whereas the partition function on spike blocks $\bloc{n_1}{n_2}$ is $Z_{n_2 - n_1} = \sum_{\bloc{n_1}{n_2}} e^{\H\bra{\bloc{n_1}{n_2}}} = \Zb^{n_2 -n_1}$. Equation (\ref{mu_ind}) expresses that spiking pattern occurring at different times are \textit{independent}
under the Gibbs distribution (\ref{Gibbs_weak}). This is expected: since the constraints
shaping $\mb$ take only into account spiking events occurring at the same time, we have no information on causality between spikes generation or on memory effects. The Gibbs distributions obtained when constructing constraints only with spatial events leads
to statistical models where spike patterns are renewed at each time step, without reference to the past activity. 

\sssu{Spatio-temporal constraints} \label{SMaxEntGen}

On the opposite, one expects that spike trains generation involves causal interactions between
neurons and memory effects. We would therefore like to construct Gibbs distributions
taking into account information on spatio-temporal interactions between neurons and leading to a statistical model not assuming anymore that successive spikes patterns 
are independent.  Although the notion of Gibbs distribution extends to processes with infinite memory \cite{fernandez-maillard:05} we shall concentrate here to Gibbs distributions associated with Markov processes with finite memory depth $D$. That is, the probability to have a spike pattern $\omega(n)$ at time $n$, given the past history of spikes reads $\Probc{\omega(n)}{\bloc{n-D}{n-1}}$. Note that those
transition probabilities are assumed not to depend explicitly on time (stationarity assumption). 

Such a family of transition probabilities $\Probc{\omega(n)}{\bloc{n-D}{n-1}}$  define an homogeneous Markov chain. Provided\footnote{This is a sufficient but not a necessary condition. In the remaining of the paper we shall work with this assumption.} $\Probc{\omega(n)}{\bloc{n-D}{n-1}} >0$ for
all $\bloc{n-D}{n}$, there is a unique probability $\mu$, called the \textit{invariant probability} of the Markov chain such that:
\beq\label{mu_inv}
\moy{\bloc{1}{D}} = \sum_{\bloc{0}{D-1}}  \Probc{\omega(D)}{\bloc{0}{D-1}} \, \moy{\bloc{0}{D-1}}.
\eeq 
In a Markov process the probability of a block $\bloc{n_1}{n_2}$,  for $n_2 -n_1+1 > D$, is:
\beq\label{ChapKol}
\moy{\bloc{n_1}{n_2}} 
= \prod_{n=n_1+D}^{n_2} \Probc{\omega(n)}{\bloc{n-D}{n-1}} \, \moy{\bloc{n_1}{n_1+D-1}},
\eeq
the Chapman-Kolmogorov relation \cite{gikhman-skorokhod:79}. To determine the probability of $\bloc{n_1}{n_2}$, one has to know the transition probabilities and the probability $\moy{\bloc{n_1}{n_1+D-1}}$. When attempting to construct a Gibbs distribution obeying (\ref{ChapKol}) from a set of spatio-temporal constraints one has therefore to determine simultaneously the family of transition probabilities and the invariant probability.
Remark that setting:
\beq\label{phi}
\phi(\bloc{0}{D}) = \log \Probc{\omega(D)}{\bloc{0}{D-1}},
\eeq
we may write (\ref{ChapKol}) in the form:
\beq\label{muGibbs}
\moyc{\bloc{n_1}{n_2}}{\bloc{n_1}{n_1+D-1}}  = e^{\sum_{n=n_1+D}^{n_2} \phi(\bloc{n}{n+D})}.
\eeq
The probability of observing the spike pattern $\bloc{n_1}{n_2}$ given the past $\bloc{n_1}{n_1+D-1}$ of depth $D$ has an exponential form, similar to (\ref{Gibbs_weak}). Actually, the invariant probability of a Markov chain is a Gibbs distribution in the following sense.\\

In view of (\ref{ChapKol}), probabilities must be defined whatever even if  $n_2 -n_1$ is arbitrary large. In this setting, the right objects are probabilities on infinite rasters \cite{gikhman-skorokhod:79}. Then, the \textit{entropy rate} (or Kolmogorov-Sinai entropy) of $\mu$ is:
\beq\label{Stat_Ent}
\s{\mu} \, = \, - \, \limsup_{n \to \infty} \frac{1}{n+1} \, \sum_{\bloc{0}{n}} \, \moy{\bloc{0}{n}} \, \log \moy{\bloc{0}{n}},
\eeq
where the sum holds over all possible blocks $\bloc{0}{n}$. This reduces to (\ref{Ssimple}) when $\mu$ obeys (\ref{mu_ind}). 

The MaxEnt takes now the following form. We consider a set of  $L$ spatio-temporal spike events
(monomials) whose empirical average value $\pTo{m_l}$ has been
computed. We only restrict to monomials with a range \textit{at most} equal to $R=D+1$, for some $D >0$. This provide us a set of constraints of the form (\ref{Constraints}). 
To maximize the entropy rate (\ref{Stat_Ent}) under the constraints (\ref{Constraints})
we construct a range-$R$ potential $\H=\sum_{l=1}^{L} \lambda_l m_l$.  The generalized form of the MaxEnt states that there is a unique probability measure $\mb \in \cM$ such that \cite{chazottes-keller:09}:
\beq\label{VarPrinc}
\p{\blambda}=\sup_{\nu \in \cM} \pare{\s{\nu} \, + \, \noy{\H)}}=
\s{\mb} \, + \, \mB{\H}.
\eeq
This is the extension of the variational principle (\ref{VarPrincWeak}) to Markov chains. It
selects, among all possible probability $\nu$,
\emph{a unique} probability $\mb$ which realizes the supremum. $\mb$ is called \textit{the Gibbs distribution with potential $\H$}. 

The quantity $\p{\blambda}$ is called \textit{topological pressure} or \textit{free energy density}.
For a potential of the form (\ref{Hh}) \cite{ruelle:69,keller:98}:
\beq\label{dPxl}
\frac{\partial \p{\blambda}}{\partial \lambda_l} = \mB{m_l}.
\eeq
This is the analog of (\ref{moy_weak}) which allows to tune the parameters $\lambda_l$.
Thus, $\p{\blambda}$ plays the role of $\log \Zb$ in (\ref{Gibbs_weak}). Actually,
it is equal to $\log \Zb$ when restricting to the memory less case\footnote{In statistical physics the free energy is $-kT \, \log Z$. The minus sign comes from the minus sign in the Hamiltonian.}.
$\p{\blambda}$ is strictly convex\footnote{Thanks to the assumption $\Probc{\omega(n)}{\bloc{n-1}{n-D}} >0$.} which guarantees the uniqueness of $\mb$.

Note that $\mu_{\blambda}$ \textit{has not} the form (\ref{Gibbs_weak}) for $D>0$. Indeed
a probability distribution e.g. of the form $\mu_{\blambda}(\bloc{0}{n-1}) = \frac{1}{Z_n} e^{\cH_{\blambda}(\omega_0^{n-1})}$ with:
\begin{equation}\label{Hn}
 \cH_{\blambda}(\bloc{0}{n-1}) \equiv \sum_{r=0}^{n-D-1} \cH_{\blambda}(\bloc{r}{r+D}) = \sum_{l}\lambda_l\sum_{r=0}^{n-D-1}m_l(\bloc{r}{r+D}),
\end{equation}
the potential of the block $\bloc{0}{n-1}$, and:
\beq\label{Zn}
Z_n\bra{\blambda}=\sum_{\bloc{0}{n-1}} e^{\H(\bloc{0}{n-1})},
\eeq
the "$n$-time steps" partition function does not obey the Chapman-Kolmogorov relation (\ref{ChapKol}).

However, the following holds \cite{ruelle:78,bowen:75,georgii:88,chazottes-keller:09}.

\begin{enumerate}

\item  There exist $A, B >0$ such that, for any block  $\bloc{0}{n-1}$:
\begin{equation}\label{BoundsGibbs}
  A \leq \frac{\mB{\bloc{0}{n-1}}}{e^{-(n-D)\p{\blambda}} e^{\H(\bloc{0}{n-1})}} \leq B.
\end{equation}
\item We have:
\begin{equation}\label{pres}
\p{\blambda}=\lim_{n \to \infty} \frac{1}{n} \log Z_n\bra{\blambda}.
\end{equation}
In the spatial case, $Z_n\bra{\blambda}=Z^n\bra{\blambda}$ and $\p{\blambda}=\log Z\bra{\blambda}$, whereas $A=B=1$ in (\ref{BoundsGibbs}). Although (\ref{pres}) is defined by a limit, it is possible to compute $\p{\blambda}$ as the $\log$ of the largest eigenvalue of a transition matrix constructed from $\cH_{\blambda}$ (Perron-Frobenius matrix)
\cite{vasquez-palacios-etal:12}. Unfortunately, this method does not apply numerically as soon as $NR>20$.

\end{enumerate} 
These relations are crucial for the developments made in the next section. \\

To recap, a Gibbs distribution in the sense of [\ref{VarPrinc}] is the invariant probability distribution of a Markov chain. The link between the potential $\cH_{\blambda}$ and the transition probabilities $\Probc{\omega(D)}{\bloc{0}{D-1}}$ (respectively the potential [\ref{phi}]) is given by: $\phi(\bloc{0}{D}) = \cH(\bloc{0}{D}) - \cG(\bloc{0}{D})$, where $\cG$, called a normalization function, is a function of the right eigenvector of a transition matrix built from $\cH$, and a function of $P[\blambda]$. $\cG$ reduces to $\log Z_{\blambda} = \p{\blambda}$ when $D=0$ [\ref{Hh}]. \\

To finish this section let us introduce the Kullback-Leibler divergence $d_{KL}(\nu,\mu)$ which provides a notion of similarity between 
two probabilities $\nu,\mu$. We have $d_{KL}(\nu,\mu) \geq 0$ with equality if and only if $\mu=\nu$. The Kullback-Leibler divergence
between an invariant probability $\nu \in \cM$ and the Gibbs distribution $\mb$ with potential $\H$ is given by $d_{KL} \pare{\nu,\mb}  = \p{\blambda} \, - \, \nu\bra{\H} \, - \, \s{\nu}$, \cite{chazottes-keller:09}. When $\nu=\pi_{\omega}^{(T)}$, we obtain the divergence between the ``model ($\mu_{\blambda}$)'' and the ``empirical probability ($\pi_{\omega}^{(T)}$)'':
\beq\label{dKL}
d_{KL} \pare{\pi_{\omega}^{(T)},\mb}  = \p{\blambda} \, - \, \pi_{\omega}^{(T)}\bra{\H} \, - \, \s{\pi_{\omega}^{(T)}}.
\eeq

\section{Inferring the coefficients of a potential from data}
\label{inferring}

Equations (\ref{moy_weak}) or (\ref{dPxl}) provide an analytical way to compute the coefficients of the Gibbs distribution from data. However, they require the computation
of the partition function or of the topological pressure which becomes rapidly intractable 
as the number of neurons increases. Thus, researchers have attempted to find alternative methods to compute reliably and efficiently the $\lambda_l$s. 
 An efficient method  has been introduced in \cite{dudik-phillips-etal:04} and applied to 
spike trains in \cite{broderick-dudik-etal:07}. Although these papers are restricted to Gibbs distributions of the form (\ref{Gibbs_weak}) (models without memory) we show in this section how their method can be extended to general Gibbs distributions.

\ssu{Bounding the Kullback-Leibler divergence variation}

\sssu{The spatial case}\label{Dudik}

The method developed in \cite{dudik-phillips-etal:04} by Dudik et al  is based on the so-called convex duality principle, used in mathematical optimization theory. Due the difficulty in maximizing the entropy (which is a concave function), one looks for a convex function easier to investigate. 
Dudik et al showed that, for spatially constrained Maxent distributions, finding the Gibbs distribution amounts to finding the minimum of the negative log likelihood\footnote{We have adapted \cite{dudik-phillips-etal:04} to our notations. Moreover, in our case $\pT$ corresponds to the empirical average on a raster $\omega$ whereas $\pi$ in \cite{dudik-phillips-etal:04} corresponds to an average over independent samples.}:
\begin{equation}
 L_{\pT}(\blambda) = - \pTo{\log \mb}.
 \label{Lpi}
\end{equation}

Indeed, in the spatial case, the 
Kullback-Leibler divergence between the empirical measure
$\pT$ and the Gibbs distribution at $\mb$
is:
\begin{equation}\label{dKLdef}
d_{KL}(\pT,\mb)=\pTo{\frac{\log \pT}{\log \mb}}
=\pTo{\log \pT} \, - \, \pTo{ \log \mb},
\end{equation}
so that, from (\ref{dKL}):
$$
L_{\pT}(\blambda) = \p{\blambda} - \pTo{\H},
$$
where we used $\s{\pi_{\omega}^{(T)}}= -\pi_{\omega}^{(T)} \Big[\log(\pi_{\omega}^{(T)})\Big]$.

Since $\cP$ is convex and $\pTo{\H}$ linear in $\blambda$, $L_{\pT}(\blambda)$ is convex. Its unique minimum is given by (\ref{moy_weak}).

Moreover, we have:
\begin{equation}\label{diffL}
L_{\pT}(\blambda') - L_{\pT}(\blambda) = \p{\blambda'} - \p{\blambda}
- \pTo{\Delta \H},
\end{equation}
with $\Delta \H = \Hp - \H$. From (\ref{Gibbs_weak}):
\begin{eqnarray}
\label{ratioz}
 \frac{Z\bra{\blambda'}}{Z\bra{\blambda}} &=& \frac{1}{Z\bra{\blambda}} \sum_{\omega(0)} e^{\Hp(\omega(0))}  \nonumber \\
  & = & \sum_{\omega(0)} e^{\Delta \H(\omega(0))} \mB{\omega(0))} \nonumber \\
  &=& \mB{e^{\Delta \cH_{\blambda}}},
\end{eqnarray}
and since $P[\blambda] = \log Z[\blambda]$ in the spatial case:
\begin{equation}\label{ZsurZ'}
\p{\blambda'} - \p{\blambda} =\log \mB{e^{\Delta \H}}.
\end{equation}
Therefore:
\beq\label{Dlambda}
L_{\pT}(\blambda') - L_{\pT}(\blambda) = \log \mB{e^{\Delta \H}} - \pTo{\Delta \H}.
\eeq

The idea proposed by Dudik et al is then to bound this difference by an easier-to-compute convex quantity, with the same minimum as $L_{\pT}(\blambda)$, and to reach this minimum by iterations on $\blambda$. They proposed a sequential and a parallel method. 
Let us summarize first the sequential method. The goal here is not to rewrite their paper \cite{dudik-phillips-etal:04} but to explain some crucial elements that are not directly appliable to the spatio-temporal case. \\ 

In the sequential case one updates $\blambda$ as $\blambda'=\blambda + \delta \be_l$, for some $l$, where $\be_l$ is the canonical basis vector in direction $l$, so that $\Delta \H=\delta \ml$,
and 
$$L_{\pT}(\blambda') - L_{\pT}(\blambda) = \log \mB{e^{\delta \ml}} - \delta \pTo{\ml}.$$ Using the following property:
\begin{equation}\label{trick}
e^{\delta x} \leq 1 + (e^\delta -1)x,
\end{equation}
 for $x \in [0,1]$ and since $\ml \in \{0,1\}$,
 we have:
\beq\label{Blog}
\log \mB{e^{\delta \ml}} \leq
\log\pare{1 + (e^\delta -1)\mb[\ml]}.
\eeq
This bound, proposed by Dudik et al, is remarkably clever. Indeed, it replaces the computation of the average $\mB{e^{\delta \ml}}$, which is computationally hard, by the computation of $\mB{m_l}$,
which is computationally easy.
Finally,
\begin{equation}\label{BoundLpiSeq}
L_{\pT}(\blambda') - L_{\pT}(\blambda)
\leq - \delta \pTo{\ml} + \log\pare{1 + (e^\delta -1)\mB{\m_l}}.
\end{equation}

\medskip

In the parallel case, the computation and results differ. One now updates $\blambda$ as $\blambda'=\blambda + \sum_{l=1}^L \delta_l \be_l$. Moreover, one has to renormalize the $m_l$s
in $m^{\prime}_l = \frac{m_l}{L}$ in order that eq. (\ref{trick2}) below holds. We have therefore $\Delta \H= \sum_{l=1}^L \delta_l m^{\prime}_l$.

Thus, 
$$L_{\pT}(\blambda') - L_{\pT}(\blambda) = \log \mB{e^{\sum_{l=1}^L \delta_l m^{\prime}_l}} - \sum_{l=1}^L \delta_l \pTo{m^{\prime}_l}.$$
 Using the following property \cite{collins-schapire-etal:02}:
\begin{equation}\label{trick2}
e^{ \sum_{l=1}^L \delta_l m^{\prime}_l} \leq 1 + \sum_{l=1}^L m^{\prime}_l \, \pare{e^{\delta_l} -1},
\end{equation}
 for $\delta_l \in \setR$ and $m^{\prime}_l \geq 0$, $ \sum_{l=1}^{L} m^{\prime}_l \leq 1$,
 we have:
$$
\log \mB{e^{\sum_{l=1}^L \delta_l m^{\prime}_l}} \leq
\log\pare{1 + \sum_{l=1}^L \pare{e^{\delta_l} -1} \, \mb[m^{\prime}_l]}.
$$
Since $\log(1+x) \leq x $ for $x > -1$, Dudick et al obtain:
$$
\log \mB{e^{\sum_{l=1}^L \delta_l m^{\prime}_l}} \leq
\sum_{l=1}^L \pare{e^{\delta_l} -1} \, \mb[m^{\prime}_l],
$$
provided $\sum_{l=1}^L \pare{e^{\delta_l} -1} \, \mb[m^{\prime}_l] > -1$ (this constraint  has to be checked during iterations).
Finally, using the definition of $m^{\prime}_l$:
\begin{equation}\label{BoundLpiPar}
L_{\pT}(\blambda') - L_{\pT}(\blambda)
\leq \frac{1}{L} \bra{- \sum_{l=1}^L \delta_l \pTo{m_l} + \sum_{l=1}^L \pare{e^{\delta_l} -1} \, \mb[m_l]}. 
\end{equation}

\medskip

To be complete, let us mention that Dudik et al consider the case where some error $\epsilon_l$ is allowed in the estimation of the coefficient $\lambda_l$. 
This relaxation on the parameters alleviates the overfitting. 

In this case, the bound
on the right hand side in (\ref{BoundLpiSeq}) (sequential case) becomes:
\begin{equation} \label{FlSeq}
F_l(\blambda,\delta)=
-\delta \pTo{\ml} \, + \, \log\pare{1+(e^\delta-1)\mB{\ml}} \, + \, \epsilon_l \pare{\abs{\lambda_l+\delta} - \abs{\lambda_l}}.
\end{equation} 
whereas the right hand side in (\ref{BoundLpiPar}) becomes $\sum_{l=1}^{L} G_l(\blambda,\bdelta)$ with:
\begin{equation} \label{FlPar}
G_l(\blambda,\bdelta)= \frac{1}{L} \bra{- \delta_l \pTo{m_l} + \pare{e^{\delta_l} -1} \, \mb[m_l]} +
\epsilon_l \pare{\abs{\lambda_l+\delta} - \abs{\lambda_l}},
\end{equation} 
The minimum of these functions is easy to find and one obtains, for a given $\blambda$
the variation $\bdelta$ required to lower bound the log-likelihood variation. 
 The authors have shown that both sequential and parallel method
produce a sequence $\blambda^{(k)}$ which
converges to the minimum of $L_{\pT}$ as $k  \to +\infty$. Note however that one strong condition in their convergence theorem is $\epsilon_l > 0$. This requires a sharp estimate of the error $\epsilon_l$, which cannot be solely based on the central limit theorem or on Hoeffding inequality in our case, because when the empirical average $\pi_{\omega}^{(T)}(m_l)$ is too small, the minima of $F$, computed in \cite{broderick-dudik-etal:07} may not be defined.

\sssu{Extension to the spatio-temporal case}

We now show how to extend these computations to the spatio-temporal case, provided one replaces the log-likelihood $L_{\pT}$ by the Kullback-Leibler divergence (\ref{dKL}). The main obstacle is that the Gibbs distribution 
does not have the form $\frac{e^\cH}{Z}$. We obtain thus a convex criterion to minimize Kullback-Leibler divergence variation, hence reaching it minimum, $\pT$. 

Replacing $\nu$ in eq. (\ref{dKL}) by $\pT$, the empirical measure,  one has:
\beq\label{dKLdiff}
d_{KL}(\pT,\mbp) - d_{KL}(\pT,\mb)=\p{\blambda'} \, - \, \p{\blambda} \, - \, \pTo{\Delta \H},
\eeq
because the entropy $\s{\pT}$ cancels. This is the analog of (\ref{diffL}). The main problem now is to compute $\p{\blambda'} \, - \, \p{\blambda}$.

From (\ref{BoundsGibbs}) we have:
\begin{eqnarray}
&& A e^{-(n-D)\p{\blambda}}  \sum_{\bloc{0}{n-1}} e^{\H(\bloc{0}{n-1})} e^{\Delta \H(\bloc{0}{n-1})} \nonumber \\
&& \leq \sum_{\bloc{0}{n-1}}  \mB{\bloc{0}{n-1}} e^{\Delta \H(\bloc{0}{n-1})} \nonumber \\
&& \leq B e^{-(n-D)\p{\blambda}}  \sum_{\bloc{0}{n-1}} e^{\H(\bloc{0}{n-1})} e^{\Delta \H(\bloc{0}{n-1})} \nonumber
\end{eqnarray}

so that:

\begin{eqnarray}
 &&
 \lim_{n \to \infty}
\frac{1}{n}
\bra{
\log A - (n-D)\p{\blambda} +
\log \Bigg( \sum_{\bloc{0}{n-1}} e^{\H(\bloc{0}{n-1})} e^{\Delta \H(\bloc{0}{n-1})} \Bigg)
}
\nonumber \\
&&
\leq
\lim_{n \to \infty}
\frac{1}{n} \log \Bigg( \sum_{\bloc{0}{n-1}}  \mB{\bloc{0}{n-1}}
e^{\Delta \H(\bloc{0}{n-1})} \Bigg)
\nonumber \\
&& 
\leq
\lim_{n \to \infty}
\frac{1}{n}
\bra{
\log B - (n-D)\p{\blambda} +
\log \Bigg( \sum_{\bloc{0}{n-1}} e^{\H(\bloc{0}{n-1})} e^{\Delta \H(\bloc{0}{n-1})}\Bigg)
}. \nonumber \\
\end{eqnarray}

Since $\Hp(\bloc{0}{n-1})=\H(\bloc{0}{n-1})+\Delta \H(\bloc{0}{n-1})$, from (\ref{pres}):
$$\lim_{n \to \infty}
\frac{1}{n} \log \sum_{\bloc{0}{n-1}}  e^{\H(\bloc{0}{n-1})}
e^{\Delta \H(\bloc{0}{n-1})} = \p{\blambda'}.$$
Therefore:
\begin{equation}\label{P'moinsP}
\p{\blambda'} - \p{\blambda}
=
\lim_{n \to \infty}
\frac{1}{n} \log \sum_{\bloc{0}{n-1}}  \mB{\bloc{0}{n-1}}
e^{\Delta \H(\bloc{0}{n-1})}.
\end{equation}
This is the extension of (\ref{ZsurZ'}) to
the spatio temporal case. In the spatial case it reduces to (\ref{ZsurZ'}) from (\ref{mu_ind}). This equation is obviously numerically intractable, but it has two advantages: on one hand it allows to extend the bounds (\ref{BoundLpiSeq}) (sequential case) and (\ref{BoundLpiPar}) (parallel case), and on the other hand it can be used to get a $\bdelta$-power expansion of $\p{\blambda'} - \p{\blambda}$. This last point  is used in the section \ref{STaylor}.\\

To get the analog of (\ref{BoundLpiSeq}) in the sequential case where $\Delta \H(\bloc{0}{n-1}) = \delta \sum_{r=0}^{n-D-1} \ml(\bloc{r}{r+D})$, one may still
apply (\ref{trick}) which holds provided:
\begin{equation}
\ml(\bloc{0}{n-1}) \equiv \sum_{r=0}^{n-1-D} \ml(\bloc{r}{r+D}) < 1 
\end{equation}
So, compared to the spatial we have to replace $\ml$
by $\frac{m_l}{n-D}$ in $\Delta \H(\bloc{0}{n-1})$. We have therefore:

\begin{eqnarray}
\sum_{\bloc{0}{n-1}} \mb[\bloc{0}{n-1}] e^{\Delta \cH(\bloc{0}{n-1})} &=& \sum_{\bloc{0}{n-1}}  \mB{\bloc{0}{n-1}} e^{\delta \frac{1}{n-D}\ml(\bloc{0}{n-1})} \nonumber \\
& \leq & 1+ (e^\delta -1) \frac{1}{n-D} \sum_{\bloc{0}{n-1}}  \mB{\bloc{0}{n-1}} \ml(\bloc{0}{n-1}). \nonumber
\end{eqnarray}

From the time translation invariance of $\mb$ we have:
\begin{eqnarray}
 \frac{1}{n-D} \sum_{\bloc{0}{n-1}} \mb[\bloc{0}{n-1}] \ml(\bloc{0}{n-1}) & = & \frac{1}{n-D} \sum_{r=0}^{n-D-1} \sum_{\bloc{0}{n-1}}\mb[\bloc{0}{n-1}]\ml(\bloc{r}{r+D}) \nonumber \\
  &=&  \frac{1}{n-D} \sum_{r=0}^{n-D-1} \mb[\ml] \nonumber \\
  &=& \mb[\ml] \nonumber
\end{eqnarray}
so that:
$$
\sum_{\bloc{0}{n-1}}  \mB{\bloc{0}{n-1}}
e^{\delta \frac{1}{n-D}\ml(\bloc{0}{n-1})} \leq 
 1+ (e^\delta -1)\mB{\ml}.
$$

At first glance this bound is not really useful. Indeed, from (\ref{P'moinsP}) we obtain:
$$
\p{\blambda'} - \p{\blambda}
\leq
\lim_{n \to \infty}
\frac{1}{n} \log \pare{1+ (e^\delta -1)\mB{\ml}} =0.
$$
Since this holds for any $\delta$ this implies $\p{\blambda'} = \p{\blambda}$.
The reason for this is evident. Renormalizing $m_l$ as we did to match 
the condition imposed by bound (\ref{trick}) is equivalent to renormalizing
$\delta$ by $\frac{\delta}{n-D}$. As $n \to +\infty$ this perturbation tends to $0$ and
 $\blambda' = \blambda$. Therefore, the clever bound (\ref{trick}) would here be of no interest
 if we were seeking exact results. However, the goal here is to propose a numerical scheme,
 where, obvioulsy $n$ is finite. We replace therefore the limit $n \to +\infty$ by a fixed $n$
 in the computation of $\p{\blambda'} - \p{\blambda}$. Keeping in mind that $\ml$
 must also be renormalized in $\pTo{\Delta \H}$ and using $\frac{1}{n} < \frac{1}{n-D}$ the Kullback-Leibler divergence (\ref{dKLdiff}) obeys:
\begin{equation}\label{BoundDKLSeq}
 d_{KL}(\pi_{\omega}^{(T)},\mbp)  -  d_{KL}(\pi_{\omega}^{(T)},\mb) \leq \frac{1}{n-D} \bra{- \delta \pTo{\ml} + \log\pare{1 + (e^\delta -1)\mB{\ml}}},
\end{equation}
the analog of (\ref{BoundLpiSeq}). \\

In the parallel case, similar remarks holds. In order to apply the bound (\ref{trick2})
we have to renormalize the $m_l$s in $m^{\prime}_l=\frac{1}{L (n-D)}$. As for the spatial case we also need to check that $\sum_{l=1}^L \pare{e^{\delta_l} -1} \, \mb[m^{\prime}_l] > -1$.
(This constraint is not guarantee and has to be checked during iterations). One obtains finally:
\begin{equation}\label{BoundDKLPar}
 d_{KL}(\pi_{\omega}^{(T)},\mbp) -  d_{KL}(\pi_{\omega}^{(T)},\mb)
\leq \frac{1}{L (n-D)} \bra{- \sum_{l=1}^L \delta_l \pTo{m_l} + \sum_{l=1}^L \pare{e^{\delta_l} -1} \, \mb[m_l]}, 
\end{equation}
the analog of (\ref{BoundLpiPar}).\\

Compared with the spatial case, we see therefore that $n$ mustn't be too large to have a reasonable  Kullback-Leibler divergence variation. It mustn't be too small, however, to get a good approximation of the empirical averages.

\ssu{Updating the target distribution when the parameters change}

When updating the parameters $\blambda$, one has to compute again the average values $\mB{\ml}$ since the probability $\mb$ has changed. This has a huge computational cost. The exact computation (e.g. from  (\ref{moy_weak}, \ref{dPxl})) is not tractable for large $N$ so approximate methods have to be used, like Montecarlo \cite{nasser-marre-etal:13}. Again, this is also CPU time consuming especially if one recomputes it again at each iteration, but at least it is tractable.

In this spirit, Broderick et al \cite{broderick-dudik-etal:07} propose to generate a Montecarlo raster distributed according to $\mb$ and to use it to compute $\mbp$ when $\|\blambda' - \blambda \|$ is sufficiently small. We explain their method, limited to the spatial case, in the next section, and we explain why it is not applicable in the spatio-temporal case. We then propose an alternative method.

\sssu{The spatial case}

The average of $\ml$ is obtained by the derivative of the topological pressure $\p{\blambda}$. In the spatial case, where $\cP(\blambda)=\log\Zb$, we have:
\begin{eqnarray}
 \mu_{\blambda^{\prime}}\bra{\ml} &=&  \frac{\partial \cP(\blambda^{\prime})}{\partial \lambda'_j} \nonumber \\
    &=&\frac{1}{Z\bra{\blambda'}} \, \sum_{\omega(0)} \ml(\omega(0)) e^{\Hp(\omega(0))} \nonumber \\
    &=&\frac{Z\bra{\blambda}}{Z\bra{\blambda'}} \, \sum_{\omega(0)} \ml(\omega(0)) e^{\Delta \H(\omega(0))} \mB{\omega(0))}
\end{eqnarray}
 Using (\ref{ratioz}),
one finally obtains:
\beq\label{Broderick}
 \mbp\bra{\ml}=\frac{\mB{\ml(\omega(0)) \, e^{\Delta \H(\omega(0))}}}{\mB{e^{\Delta \H(\omega(0))}}},
\eeq
which is eq. (18) in \cite{broderick-dudik-etal:07}. Using this formula one is able to compute the average of $\ml$ with respect to the new probability $\mbp$ only using the old one, $\mb$. 

\sssu{Extension to the spatio-temporal case}

We now explain why the Broderick et al method does not extend to the spatio-temporal case. 
The main problem is that if one tries to obtain the analog of the equality (\ref{Broderick})
one obtains in fact an inequality:
\beq\label{IBrod}
\frac{A}{B}\mu_{\blambda^{\prime}}\bra{\ml}
\leq
\lim_{n \to \infty} \frac{1}{n} 
\frac{\mB{\ml\pare{\bloc{0}{n-1}} e^{\Delta \H(\bloc{0}{n-1})}}}{\mB{e^{\Delta \H(\bloc{0}{n-1})}}}
\leq 
\frac{B}{A}\mu_{\blambda^{\prime}}\bra{\ml},
\eeq
where $A,B$ are the constants in (\ref{BoundsGibbs}). They are not known in general (they depend on the potential) and they are different. However, in the spatial case $A=B=1$
whereas $\mB{\ml\pare{\bloc{0}{n-1}} e^{\Delta \H(\bloc{0}{n-1})}} = \mB{\ml\pare{\omega(0)} e^{\Delta \H(\omega(0))}}$ because the potential has range $1$. Then, one recovers (\ref{Broderick}). Let us now explain how we obtain (\ref{IBrod}).\\

The averages of quantities are obtained by the derivative of the topological pressure (Eq. (\ref{dPxl})).
 We have:
\begin{equation}\label{DP}
\mu_{\blambda^{\prime}}\bra{\ml}=
 \frac{\partial \cP}{\partial \lambda'_l}=
\frac{\partial \lim_{n \to \infty}  \frac{1}{n} \log Z_n\bra{\blambda'}}{\partial \lambda'_l}.
\end{equation}

Assuming that the limit and the derivative commute (see e.g. \cite{mayer-urbanski:10}), gives: 
\begin{eqnarray}
 \mu_{\blambda^{\prime}}\bra{\ml} &=& \lim_{n \to \infty} \frac{1}{n} \frac{1}{Z_n\bra{\blambda'}} \sum_{\bloc{0}{n-1}} \ml\pare{\bloc{0}{n-1}} e^{\Hp(\bloc{0}{n-1})} \nonumber \\
  &=&\lim_{n \to \infty} \frac{1}{n} \frac{1}{Z_n\bra{\blambda'}} \sum_{\bloc{0}{n-1}} \ml\pare{\bloc{0}{n-1}} e^{\Delta \H(\bloc{0}{n-1})} e^{\H(\bloc{0}{n-1})}  \nonumber \\
  &=&\lim_{n \to \infty} \frac{1}{n} \frac{\sum_{\bloc{0}{n-1}} \ml\pare{\bloc{0}{n-1}} e^{\Delta \H(\bloc{0}{n-1})} e^{\H(\bloc{0}{n-1})}}{\sum_{\bloc{0}{n-1}}  e^{\Delta \H(\bloc{0}{n-1})} e^{\H(\bloc{0}{n-1})}} \nonumber \\.
  \label{48}
\end{eqnarray}

From (\ref{BoundsGibbs}):

\begin{eqnarray}
  && A \, e^{-(n-D)\p{\blambda}}  \sum_{\bloc{0}{n-1}} \ml\pare{\bloc{0}{n-1}} e^{\Delta \H(\bloc{0}{n-1})} 
e^{\H(\bloc{0}{n-1})} \nonumber \\
 && \leq    \sum_{\bloc{0}{n-1}} \ml\pare{\bloc{0}{n-1}} e^{\Delta \H(\bloc{0}{n-1})} 
\mB{\bloc{0}{n-1}} \nonumber \\
&&\leq B \, e^{-(n-D)\p{\blambda}}   \sum_{\bloc{0}{n-1}} \ml\pare{\bloc{0}{n-1}} e^{\Delta \H(\bloc{0}{n-1})} 
e^{\H(\bloc{0}{n-1})} \nonumber \\
\end{eqnarray}

and:
\begin{eqnarray}
&& A \, e^{-(n-D)\p{\blambda}}  \sum_{\bloc{0}{n-1}}  e^{\Delta \H(\bloc{0}{n-1})} e^{\H(\bloc{0}{n-1})} \nonumber \\
&& \leq    \sum_{\bloc{0}{n-1}}  e^{\Delta \H(\bloc{0}{n-1})} \mB{\bloc{0}{n-1}} \nonumber \\
&& \leq B \, e^{-(n-D)\p{\blambda}}   \sum_{\bloc{0}{n-1}}  e^{\Delta \H(\bloc{0}{n-1})} e^{\H(\bloc{0}{n-1})} \nonumber .
\end{eqnarray}

Therefore:
$$
\frac{A}{B} \frac{\sum_{\bloc{0}{n-1}} \ml\pare{\bloc{0}{n-1}} e^{\Delta \H(\bloc{0}{n-1})} 
e^{\H(\bloc{0}{n-1})}}{\sum_{\bloc{0}{n-1}}  e^{\Delta \H(\bloc{0}{n-1})} 
e^{\H(\bloc{0}{n-1})}}
$$
$$
\leq
\frac{\sum_{\bloc{0}{n-1}} \ml\pare{\bloc{0}{n-1}} e^{\Delta \H(\bloc{0}{n-1})} 
\mB{\bloc{0}{n-1}}}{\sum_{\bloc{0}{n-1}}  e^{\Delta \H(\bloc{0}{n-1})} 
\mB{\bloc{0}{n-1}}}
$$
$$
\leq \frac{B}{A} \frac{\sum_{\bloc{0}{n-1}} \ml\pare{\bloc{0}{n-1}} e^{\Delta \H(\bloc{0}{n-1})} 
e^{\H(\bloc{0}{n-1})}}{\sum_{\bloc{0}{n-1}}  e^{\Delta \H(\bloc{0}{n-1})} 
e^{\H(\bloc{0}{n-1})}}.
$$
Now, from \cite{chazottes-keller:09,keller:98}, (\ref{48}) gives (\ref{IBrod}).


\sssu{Taylor expansion of the pressure}\label{STaylor}

The idea is here to use a Taylor expansion of  the topological pressure. This approach is very much in the spirit of \cite{kappen-rodriguez:98}, but extended here to the spatio-temporal case.
Since $\blambda'=\blambda+\bdelta$, we have:
\begin{eqnarray}
 \mu_{\blambda^{\prime}}\bra{\ml} &=& \mB{\ml} + \sum_{j=1}^L 
\frac{\partial \mB{\ml}}{\partial \lambda_j} \delta_j + \frac{1}{2} \sum_{j,k=1}^L 
\frac{\partial^2 \mB{\ml}}{\partial \lambda_j \partial \lambda_k} \delta_j \delta_k + \dots \nonumber \\
 &=& \mB{\ml} + \sum_{j=1}^L 
\frac{\partial^2 \p{\blambda}}{\partial \lambda_j \partial \lambda_l} \delta_j + \frac{1}{2} \sum_{j,k=1}^L 
\frac{\partial^3 \p{\blambda}}{\partial \lambda_j \partial \lambda_k \partial \lambda_l} \delta_j \delta_k + \dots
\end{eqnarray}

The second derivative of the pressure is given by \cite{ruelle:78,bowen:75,georgii:88,chazottes-keller:09}:
\begin{equation}\label{FDT}
\frac{\partial^2 \p{\blambda}}{\partial \lambda_j \partial \lambda_l} =
\sum_{n=-\infty}^{+\infty} C_{jl}(n) \equiv \chi_{jl}\bra{\blambda},
\end{equation}
where:
\begin{equation}
C_{jl}(n)=\mB{m_j \, \ml \circ \sigma^n} - \mB{m_j} \mB{\ml},
\end{equation}
is the correlation function between $\ml, \mk$ at time $n$, computed with respect to $\mb$. (\ref{FDT}) is a version of the fluctuation-dissipation theorem in the spatio-temporal case.
$\sigma^n$ is the time shift applied $n$ times.
 The third derivatives can be computed as well by taking the derivative (\ref{FDT}) and using (\ref{DP}). This generates terms with third order correlations and so on \cite{mayer-urbanski:10}. Up to second order we have:

\begin{equation}\label{Taylor}
\mu_{\blambda^{\prime}}\bra{\ml} = \mB{\ml} + \sum_{j=1}^L 
\chi_{jl}\bra{\blambda} \delta_j + \dots
\end{equation}

Since the observable are monomials they only take the values 0 or 1 and the computation of $\chi_{jl}$
is straightforward, reducing to counting the occurrence of time pairs $t,t+n$ such that $m_j(t)=1$ and $\ml(t+n)=1$.\\
 
On practical grounds we introduce a parameter $\Delta=\|\blambda' - \blambda \| $ which measures the variation in the parameters after update. If $\Delta$ is small enough (smaller than some $\Delta_c$), the terms of order $3$ in the Tayor expansion are negligible, then we can use (\ref{Taylor}). Otherwise, if $\Delta$ is big, we compute a new Montecarlo estimation of $\mb'$ (as described in \cite{nasser-marre-etal:13}). We explain in section \ref{SDeltac} how $\Delta_c$ was chosen in our data.   Then, we use the following trick.  If $\|\bdelta\| > \Delta_c$ we compute the new value $\mu_{\blambda^{\prime}}\bra{\mj}$. If $\Delta_c > \|\bdelta\| > \frac{\Delta_c}{10}$, we use the linear response approximation (\ref{Taylor}) of $\mu_{\blambda^{\prime}}$. Finally, if $\|\bdelta\| < \frac{\Delta_c}{10}$ we use $\mB{\ml}$ instead of $\mu_{\blambda^{\prime}}\bra{\ml}$ in the next iteration of the method . Thus, in the case, $\|\bdelta\| < \Delta_c$, we use the Gibbs distribution computed at some time step, say $n$, 
to infer the values at the next iteration. If we do that several successive time steps the distance to the original value $\blambda_n$ of the parameters increases. So we compute the norm $\| \blambda_n - \blambda_{n+k} \|$ at each time step $k$, and we do not compute a new raster until this norm is larger than $\Delta_c$.

\ssu{The algorithms}\label{Salgo}

We have two algorithms, sequential and parallel, which are very similar to Dudik el al. Especially, the convergence of their algorithms, proved in their paper, 
extends to our case since it only depends on the shape of the cost functions
(\ref{FlSeq}, \ref{FlPar}).
We describe here the algorithms coming out from the presented mathematical framework, in a sequential and parallel version. 
We iterate the algorithms until
the distance $\eta = d\pare{\mu_{\blambda},\pi_{\omega}^{(T)}}$
is smaller than some $\eta_c$. We use the Hellinger distance:

\beq\label{Hellinger}
d\pare{\mu_{\blambda},\pi_{\omega}^{(T)}} = \frac{1}{\sqrt{2}} \sqrt{\sum_{l=1}^L \pare{\sqrt{\pi_{\omega}^{(T)}(\ml)} - \sqrt{\mu_{\blambda}(\ml)}}^2}
\eeq


\sssu{Sequential algorithm}


\begin{algorithm}[H]
\DontPrintSemicolon
 \SetAlgoLined
 \KwIn{The features empirical probabilities $\pTo{m_l}$}
 \KwOut{ The vector of parameters $\blambda$ }
 initialization: $\lambda_l = 0$ for every $l$, $\Delta = 0$\;
 \While{$\eta > \eta_c$}{
   $(\delta,l) = \arg \min_{l, \delta} F_l(\blambda,\delta)$\;
   $\lambda_l \leftarrow \lambda_l+\delta$\;
   $\Delta \leftarrow \sqrt{\Delta^2+\delta^2}$ \;
   \eIf{$\Delta > \Delta_c$} {Compute a new Gibbs sample using Montecarlo method \cite{nasser-marre-etal:13}}
   {Compute the new features probabilities using Taylor expansion (Equation \ref{Taylor})}

 }
 \caption{Sequential algorithm. $\delta$ is the learning rate by which we change the value of a parameter $\lambda_l$. $\eta$ is the convergence criterion (\ref{Hellinger})). $\Delta$ is the parameter allowing us to decide whether we update the parameters change by computing a new Gibbs sample or by the Taylor expansion. $F_l$ is given by eq. (\ref{FlSeq})}.
\end{algorithm}

\ssu{Parallel algorithm}
%

\begin{algorithm}[H]
\DontPrintSemicolon
 \SetAlgoLined
 \KwIn{The features empirical probabilities $\pTo{m_l}$}
 \KwOut{parameters $\lambda_l$ }
 initialization: $\lambda_l = 0$ for every $l$, $\Delta = 0$\;
 \While{$\eta > \eta_c$}{
   \For{$l\leftarrow 1$ \KwTo $L$}{
   $\delta_l = \arg \min_{\delta} G_l(\blambda,\bdelta)$
   }
   $\blambda \leftarrow \blambda+\bdelta$\;
   $\Delta \leftarrow \sqrt{\Delta^2 + \sum_{l=1}^L \delta_l^2}$ \;
   \eIf{$\Delta > \Delta_c$} {Compute a new Gibbs sample using Montecarlo method \cite{nasser-marre-etal:13}}
   {Compute the new features probabilities using Taylor expansion (Equation \ref{Taylor})}

 }
  \caption{The parallel algorithm. $G_l$ is given by (\ref{FlPar}).}
\end{algorithm}

The implementation of those algorithms consists on an important part in a software developed at INRIA and called EnaS (Event Neural Assembly Simulation). The executable is freely available at \url{http://enas.gforge.inria.fr/v3/download.html}.

\section{Results}
\label{tests}

In this section we perform several tests on our method. We first consider synthetic data generated with a known Gibbs potential and recover its parameters. This step also allows us to tune the parameter $\Delta_c$ in the algorithms. Then, we  consider real data analysis where the Gibbs potential form is unknown. This last step is not a systematic study that would be out of the scope of this paper, but simply provided as an illustration and comparison with the paper of Schneidman et al. 2006 \cite{schneidman-berry-etal:06}.

\subsection{Synthetic data}
\label{synth}

Synthetic data are obtained by generating a raster distributed according to a Gibbs distribution whose potential (\ref{Hh}) is known. 
We consider two families of Gibbs potentials. For each family there are $L > N$ monomials whose range belongs to $\Set{1, \dots , R}$. Among them, there are $N$ "rate monomials" $\omega_i(D), \, i=1 \dots N$, whose average  gives the firing rate of neuron $i$, denoted $r_i$ ; the $L-N$ other monomials, with degree $k>1$,  are chosen at random with a probability law $\sim e^{-k}$ which favors therefore pairwise interactions. The difference between the two families comes from the distribution of coefficients $\lambda_l$.
 
\begin{enumerate}
\item  \textbf{"Dense" rasters family}. 
 The coefficients are drawn with a Gaussian distribution with mean $0$ and variance $\frac{1}{L}$ to ensure a correct scaling of the coefficients dispersion as $L$ increases (Figure \ref{Fdensea}). This produces typically a dense raster (Figure \ref{Fdenseb})  with strong multiple correlations.
 \begin{figure}[ht!]
\centering
\subfigure[Example of coefficients distribution in the dense rasters family.]{
 \centering
 \includegraphics[width=12cm,height=6cm]{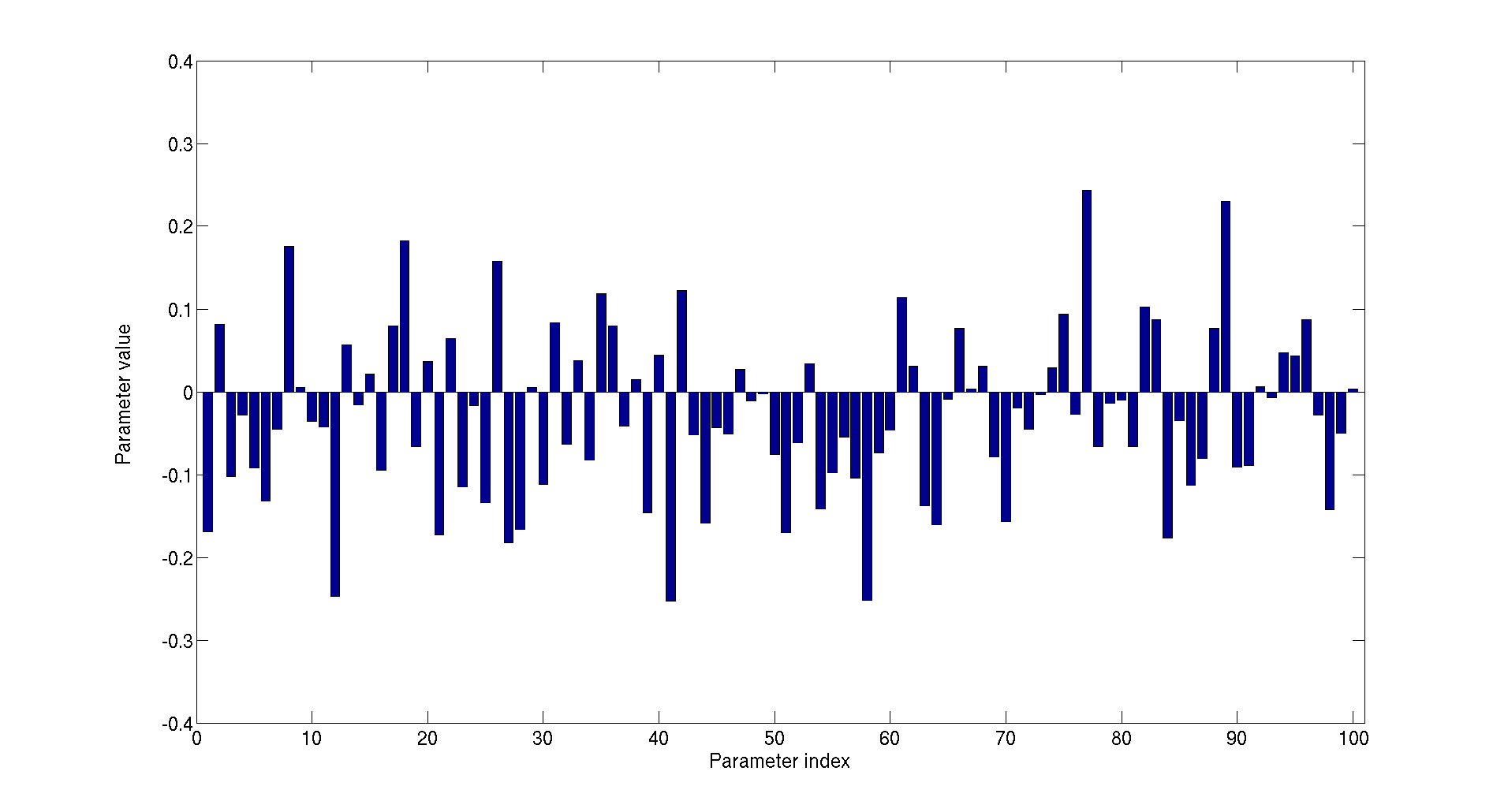}
 \label{Fdensea}
}
\subfigure[Dense spike train]{
 \centering
 \includegraphics[width=10.5cm,height=5cm]{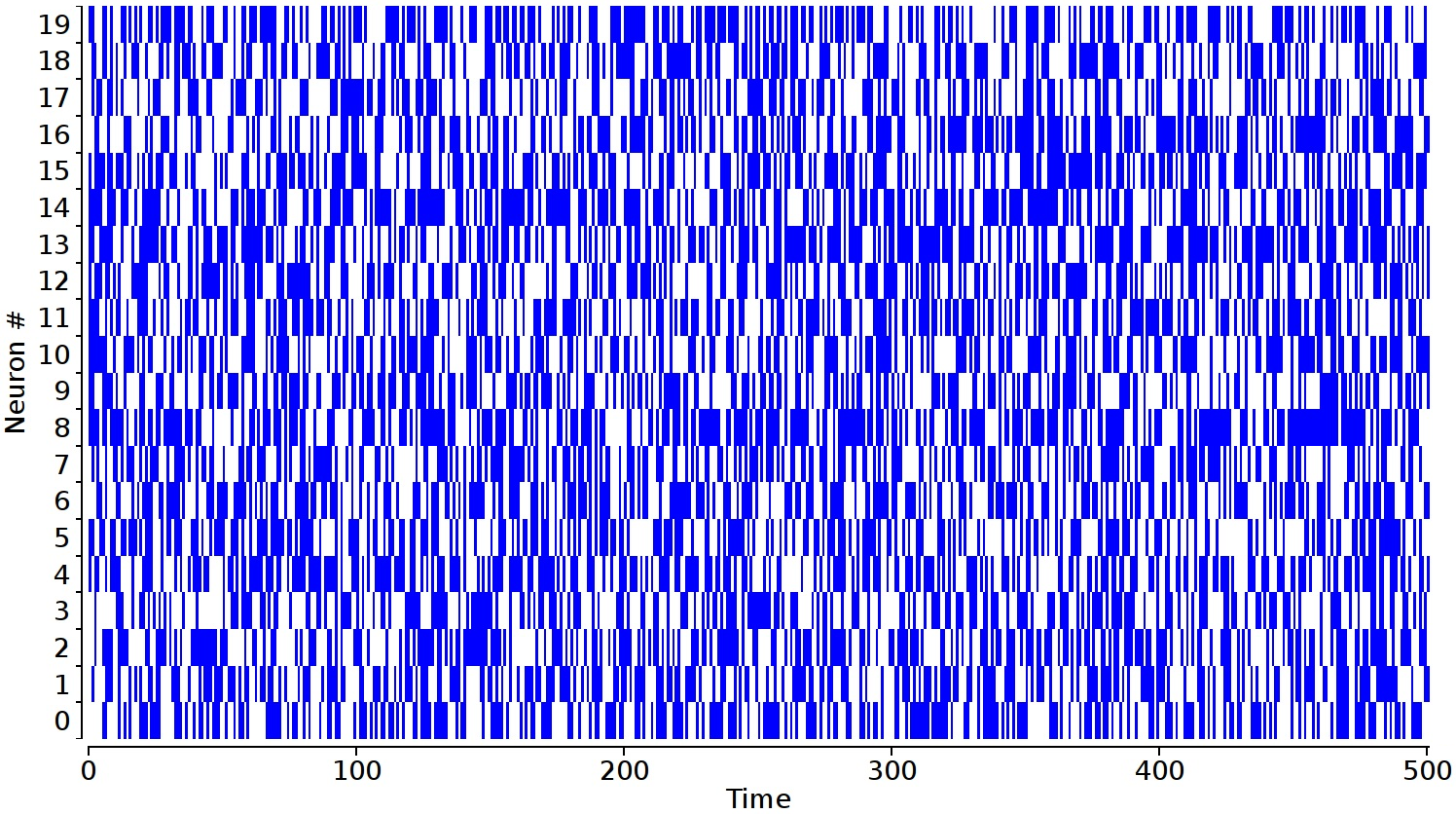}
 \label{Fdenseb}
}

\caption{Dense family.}

\end{figure}

\item \textbf{"Sparse" rasters family.} 
The rate coefficients in the potential are very negative: the coefficient $h_i$
of the rate monomial $\omega_i(D)$ is $h_i = \log\pare{\frac{r_i}{1-r_i}}$ where $r_i \in [0:0.01]$ with a uniform probability distribution. Other coefficients are drawn with a Gaussian distribution with mean $0.8$ and variance $1$ (Figure \ref{Fsparsea}). This produces a sparse raster (Figure \ref{Fsparseb}) with strong multiple correlations. 

 \begin{figure}[ht!]

\centering
\subfigure[Example of coefficients distribution in the sparse rasters family.]{
 \centering
 \includegraphics[width=12cm,height=6cm]{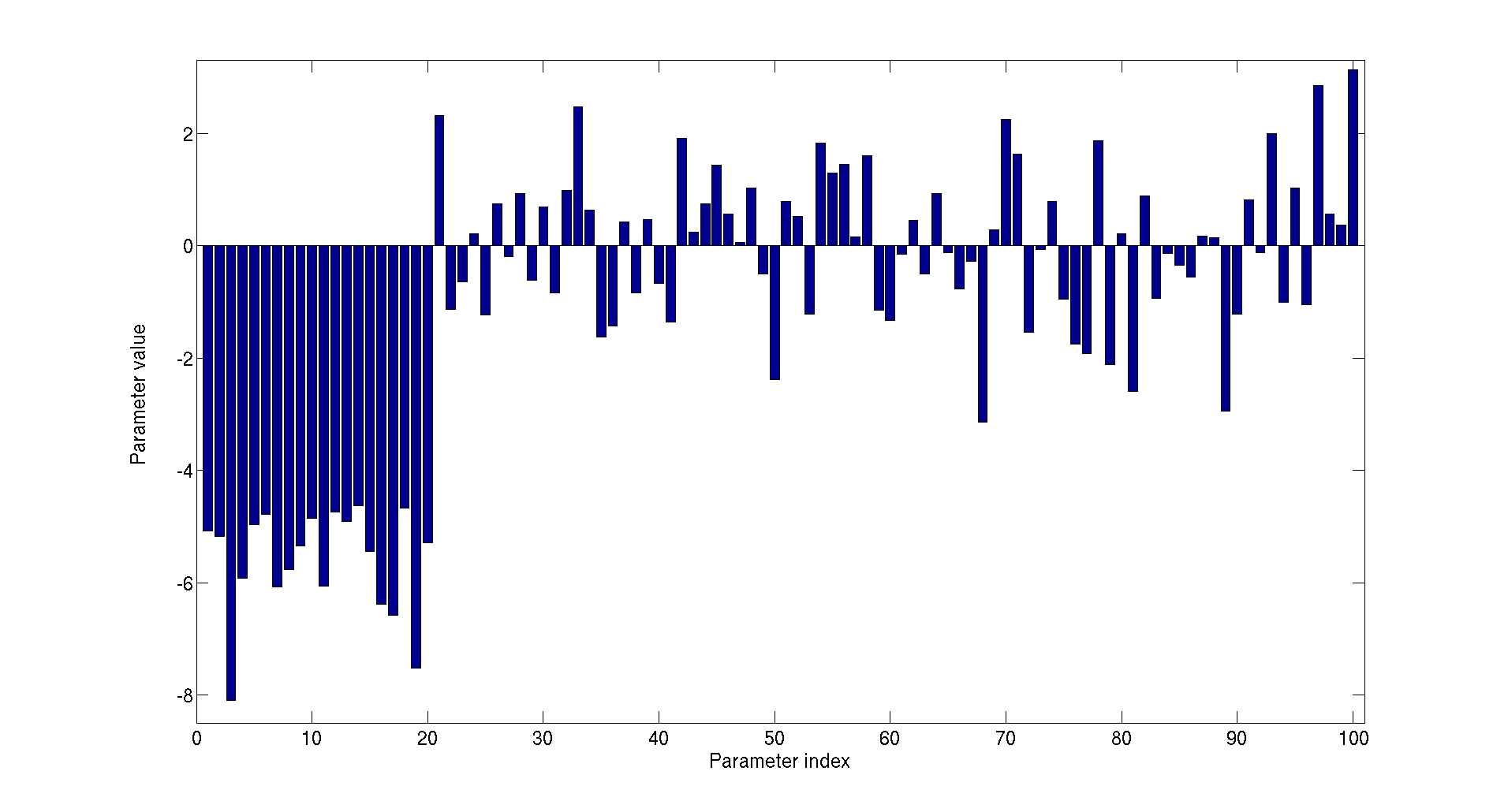}
\label{Fsparsea}
}
\subfigure[Sparse spike train]{
 \centering
 \includegraphics[width=10.5cm,height=5cm]{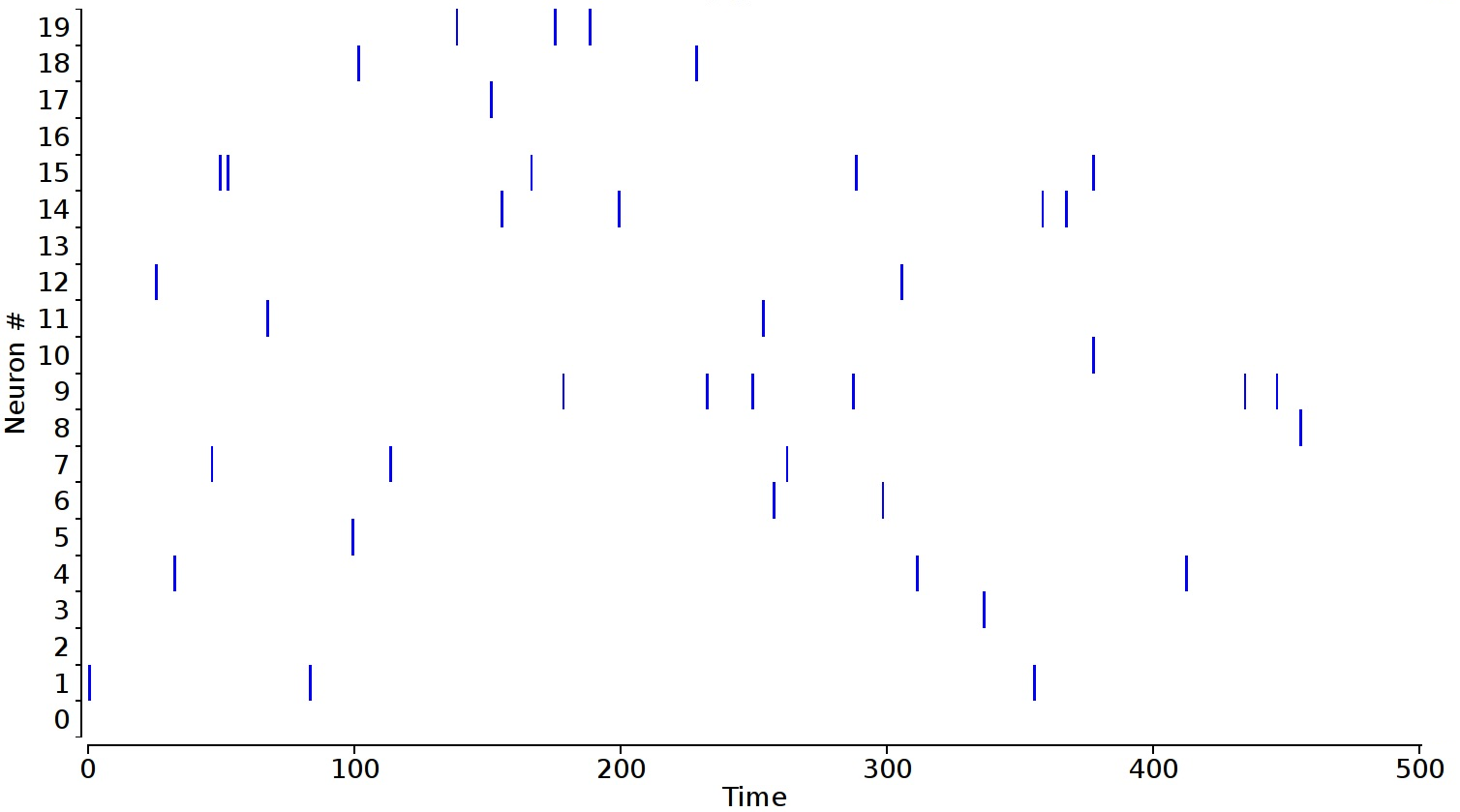}
\label{Fsparseb}
}
\caption{Sparse family.}
\end{figure}
\end{enumerate}

\pagebreak

\subsection{Tuning $\Delta_c$}\label{SDeltac}
For small $N,R$ ($NR \leq 20$) it is possible to exactly compute the topological pressure
using the transfer matrix technique \cite{vasquez-marre-etal:12}. We have therefore a way to compare the Taylor expansion (\ref{FDT}) and the exact value.

If we perturb $\blambda$ by an amount $\delta$ in the direction $l$,
 this induces a variation on $\mB{\ml}$, $l=1 \dots L$, given by the Taylor expansion (\ref{Taylor}). To the lowest order $\mu_{\blambda^{\prime}}\bra{\ml} = \mB{\ml} + O^{(1)}$,
 so that:
$$
\epsilon^{(1)} = \frac{1}{L} \, \sum_{l=1}^L \frac{\abs{\mu_{\blambda^{\prime}}\bra{\ml} -\mB{\ml}}}{\abs{\mu_{\blambda^{\prime}}\bra{\ml}}}
$$
 is a measure of the relative error when considering the lowest order expansion. 
 
In the same way, to the second order:
$$
\mu_{\blambda^{\prime}}\bra{\ml} = \mB{\ml} +  \sum_{j=1}^L\chi_{jl}\bra{\blambda} \delta_j + O^{(2)},
$$
so that:
$$
\epsilon^{(2)} = \frac{1}{L} \, \sum_{l=1}^L \frac{\abs{\mu_{\blambda^{\prime}}\bra{\ml} -\mB{\ml} - \sum_{j=1}^L\chi_{jl}\bra{\blambda} \delta_j}}{\abs{\mu_{\blambda^{\prime}}\bra{\ml}}},
$$
 is a measure of the relative error when considering the next order expansion.

\begin{figure}[htbp]
\begin{center}
\includegraphics[height=5.5cm,width=5.5cm]{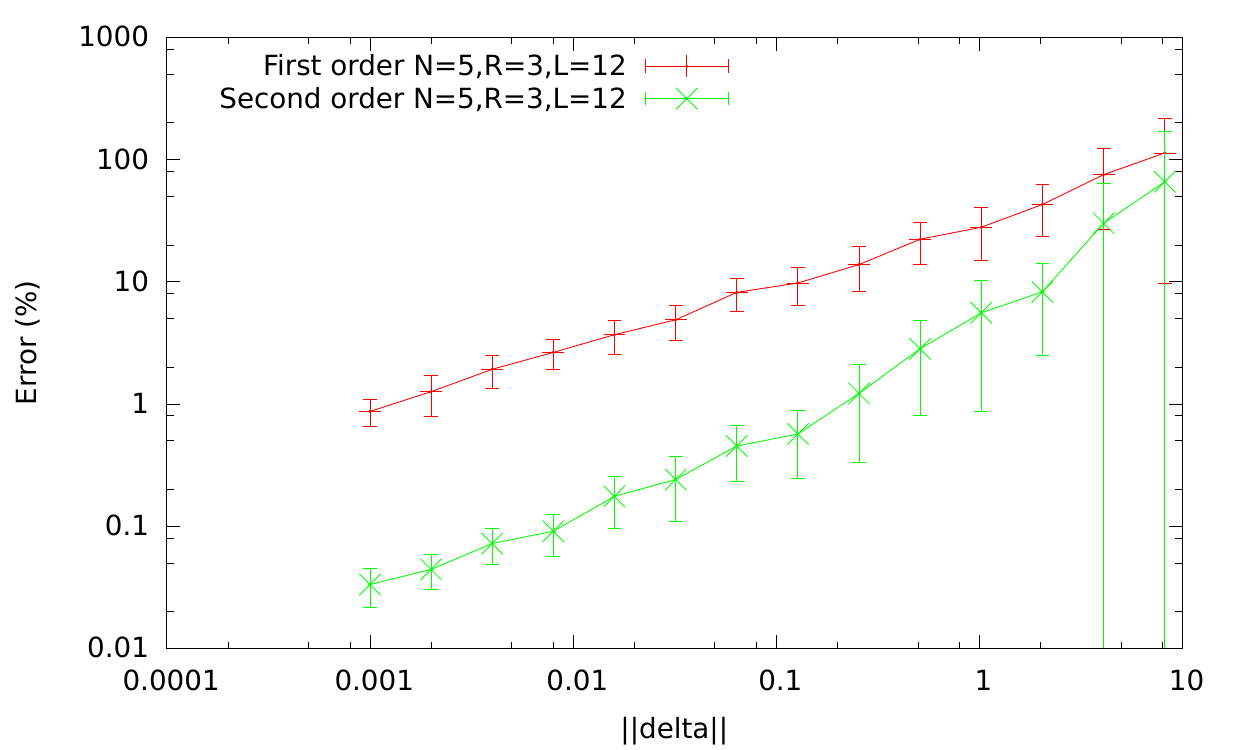}
\hspace{0.5cm} 
\includegraphics[height=5.5cm,width=5.5cm]{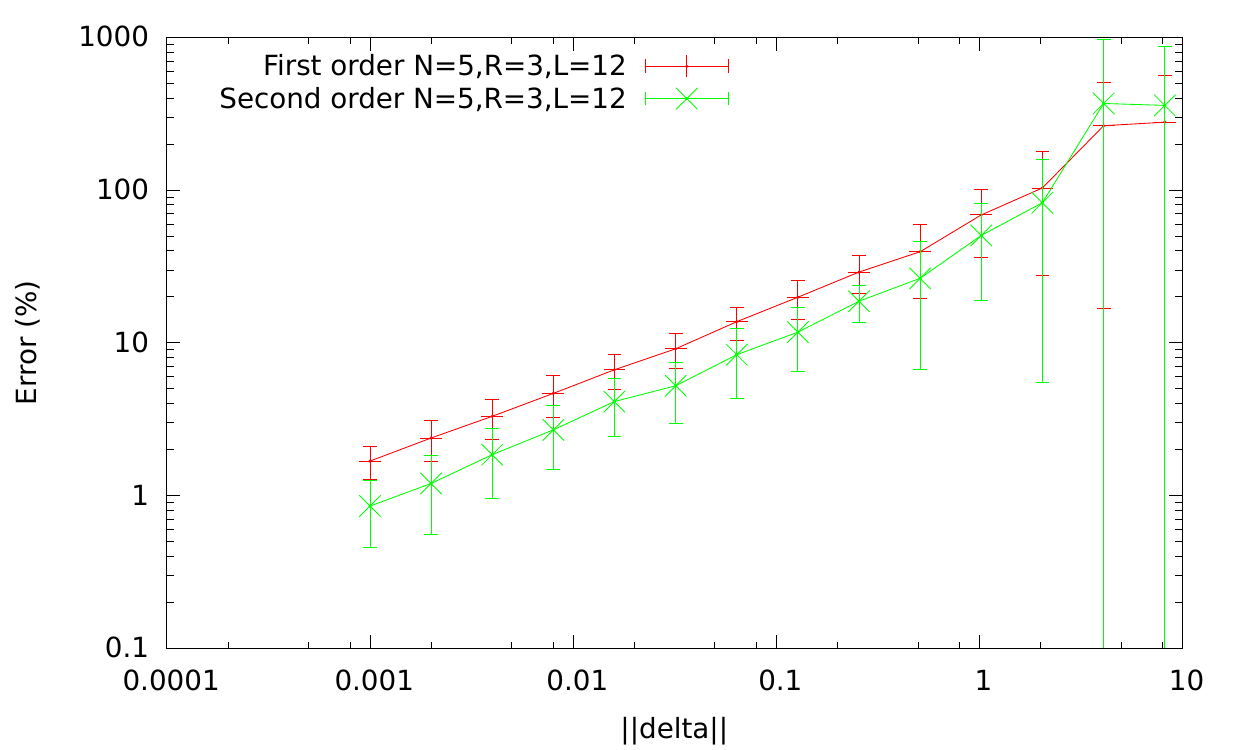}
\caption{Error on the average $\mu_{\blambda^{\prime}}\bra{\ml}$ as a function of the perturbation amplitude $\delta$. First order corresponds to $\epsilon^{(1)}$ and second order to $\epsilon^{(2)}$ (see text). The curves correspond to $N=5,R=3,L=12$. Left: Dense case;
Right: Sparse case.}
\label{FVariations}
\end{center}
\end{figure}

In Figure \ref{FVariations} we show the relative errors $\epsilon^{(1)},\epsilon^{(2)}$ (in $\%$), as a function of $\delta$. For each point we generate  $25$ potentials, with $N=5,R=3,L=12$. For each of these potentials we randomly perturb the $\lambda_j$s, with a random sign, so that the norm of the perturbation $\|\bdelta\|$ is fixed. The linear response $\chi$ is computed from a raster of length $T=100000$.

These curves show a big difference between the  dense and sparse case.
In the dense case, the second order error is about $5 \%$ for $\Delta_c=1$
whereas we need a $\Delta_c \sim 0.03$ to get the same $5 \%$ in the sparse case. We choose to align
on the sparse case and in typical experiments we take $\Delta_c=0.1$ corresponding
to about $10 \%$ of error on the second order.

\subsection{Computation of the Kullback-Leibler divergence}

To compute the Kullback-Leibler divergence between the empirical distribution $\pi_{\omega}^{(T)}$ and the fitted predicted distribution $\mu_{\blambda}$ , we need to know the value of the pressure $\p{\blambda}$, the empirical probability of the potential $\pi_{\omega}^{(T)}\bra{\H}$ and the entropy $\s{\pT}$. For small networks, we can compute the pressure using the Perron-Frobenius theorem (\cite{vasquez-marre-etal:12}). However, for large scales, since we cannot compute the pressure, computing the Kullback-Leibler divergence is not direct and exact. We compute an approximation using the following technique. From Eq. (\ref{VarPrinc}) and (\ref{dKL}), we can write:

\begin{eqnarray}\label{dkl_cool}
 d_{kl}(\pT, \mb) & = & \mb\bra{\H} + \s{\mb} - \pT\bra{\H} - \s{\pT} \\ \nonumber
   & = & \sum_l \lambda_l \big(\mb[m_l] - \pTo{m_l}\big)  + \s{\mb} - \s{\pT}
\end{eqnarray}

From the parameters $\blambda$, we compute a spike train distributed as $\mu_{\blambda}$ using the Montecarlo method (\cite{nasser-marre-etal:13}). From this spike train, we compute the monomials averages $\mb[m_l]$ and the entropy $\s{\mb}$ using the method of Strong et al. (\cite{strong-koberle-etal:98}). $\pTo{m_l}$ and $\cS[\pT]$ are computed directly on the empirical data set.

\subsection{Performances on synthetic data}\label{Ssynth}

Here, we test the method on synthetic data where the shape of the sought potential is known: only the $\lambda_l$s have to be estimated. Experiments were designed according to the following steps:

\begin{itemize}
 \item We start from a potential $\Ha = \sum_{l \in \cL} \lambda^\ast_l m_l$. The goal is to estimate the coefficient values $\lambda^\ast_l$ knowing the set $\cL$ of monomials spanning the potential.
 \item We generate a synthetic spike train ($\omega_{s}$) distributed according to the Gibbs distribution of $\Ha$.
 \item We take a potential $\cH_{\blambda} = \sum_{l \in \cL} \lambda_l m_l$ with random initial coefficients $\lambda_l$. Then we fit the parameters $\lambda_l$ to the synthetic spike train $\omega_{s}^{(T)}$. 
 \item We evaluate the goodness of fit. 
\end{itemize}

\begin{figure}[h]
 \centering
 \subfigure[]{
\includegraphics[height=4cm,width=5cm]{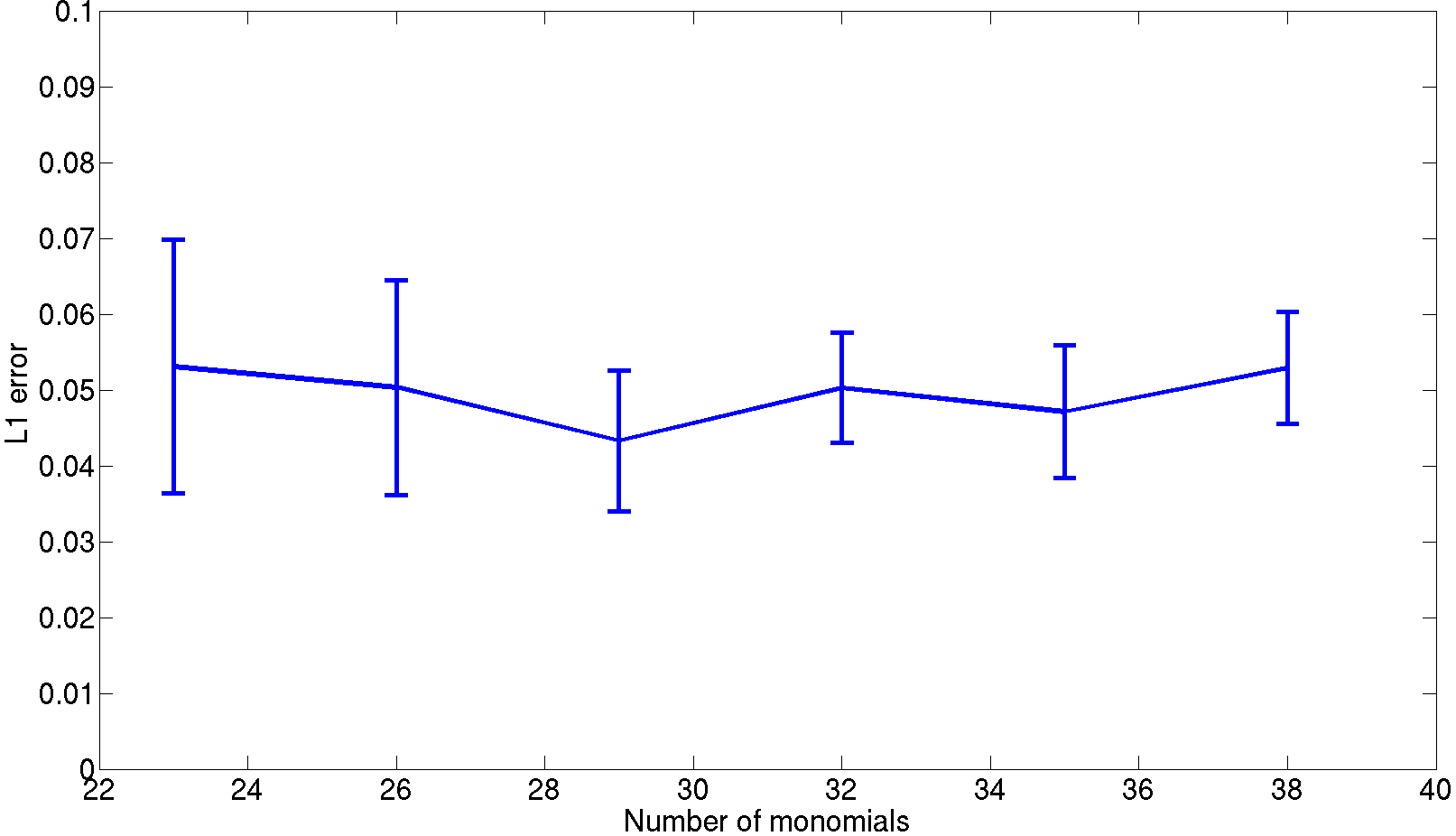}
 \label{dense}
 }
 \centering
 \subfigure[]{
\includegraphics[height=4cm,width=5cm]{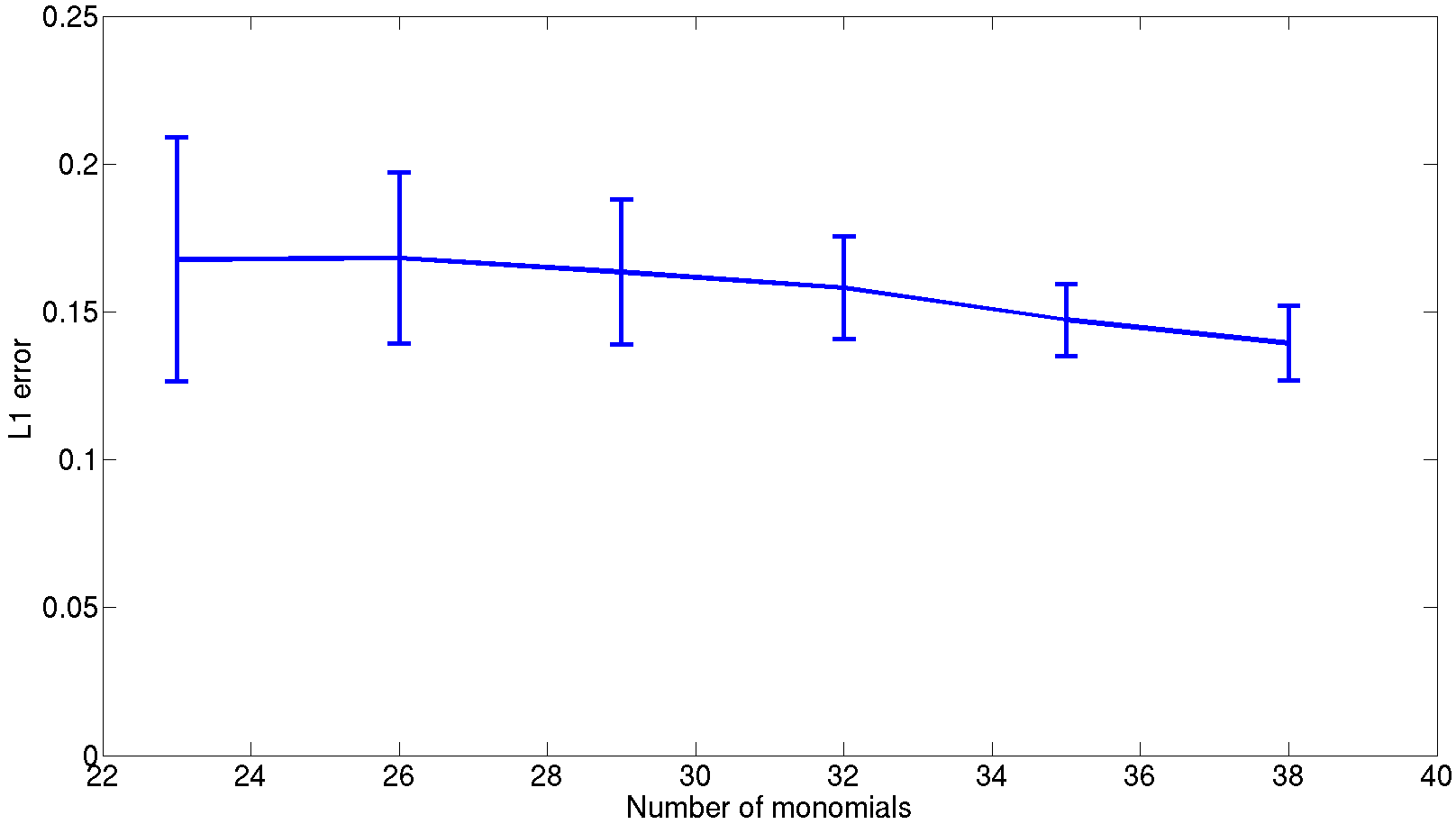}
 \label{sparse}
 }
\caption{Distance between the exact value of coefficients and the estimated value,
averaged on the set of $10$ random potentials for $NR = 60$. (a) Dense spike trains (b) Sparse spike trains.}
 \label{FLargeNR}
\end{figure}

For the last step (goodness of fit) we have used three criteria. The first one simply consists of computing the $L_1$ error $d_1 = \frac{1}{L} \sum_{l=1}^L \abs{\lambda^\ast_l - \lambda^{(est)}_l}$ where  $\lambda^{(est)}_k$ is the final estimated value. $d_1$ is then averaged on  $10$ random potentials.  Fig. \ref{FLargeNR} shows the committed error in the case of sparse and dense potentials.
The method showed a good performance, both in dense and sparse case, for large $N  \times R \sim 60$. 

The main advantage of this criterion is to provide an exact estimation of the error made on coefficients estimation. Its drawback is that we have to know the shape of the potential which generated the raster: this is not the case anymore for real neural networks data. We therefore
used a second criterion: confidence plots. For each spike block $\bloc{0}{D}$ appearing in the raster $\omega_{s}$ we draw a point in a two dimensional diagram with, on abscissa, the observed empirical probability $\pTos{\bloc{0}{D}}$ and, on ordinate, the predicted probability $\mB{\bloc{0}{D}}$. Ideally, all points should align on the diagonal $y=x$ (equality line). However, since the raster is finite there are finite-size fluctuations ruled by the central limit theorem.
For a block $\bloc{0}{D}$ generated by a Gibbs distribution $\mb$ and having an exact probability $\mB{\bloc{0}{D}}$ the empirical probability $\pTos{\bloc{0}{D}}$ is a Gaussian random variable with mean  $\mB{\bloc{0}{D}}$ and mean-square deviation $\sigma=\frac{\sqrt{\mB{\bloc{0}{D}}\pare{1- \mB{\bloc{0}{D}}}}}{\sqrt{T}}$. The probability that $\pTos{\bloc{0}{D}} \in \left[\mB{\bloc{0}{D}} - 3\sigma, \mB{\bloc{0}{D}} + 3\sigma \right]$ is therefore of about $99,6 \%$. This interval is represented by confidence lines spreading around the diagonal.  As a third criterion, we have used the Kullback-Leibler divergence (\ref{dkl_cool}).

We have plotted 2 examples in Figures \ref{Sisingsynthetic} \& \ref{Spairwisesynthetic} for sparse data types:
\begin{enumerate}
 \item Spatial case, 40 neurons, ($NR = 40$): Ising model (\ref{HIsing}).  Fig. \ref{Sisingsynthetic} .
  \begin{figure}[!]
 \centering
   \centering
    \subfigure[Monomials averages]{
\includegraphics[height=4cm,width=0.45\textwidth]{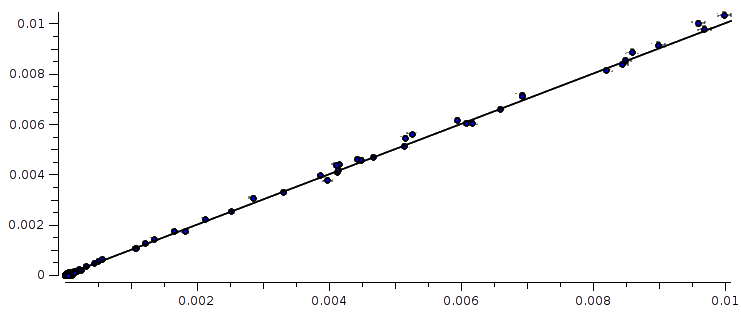}
 \label{sr2monomials}
 }
 \subfigure[Patterns of depth 1]{
\includegraphics[height=4cm,width=0.45\textwidth]{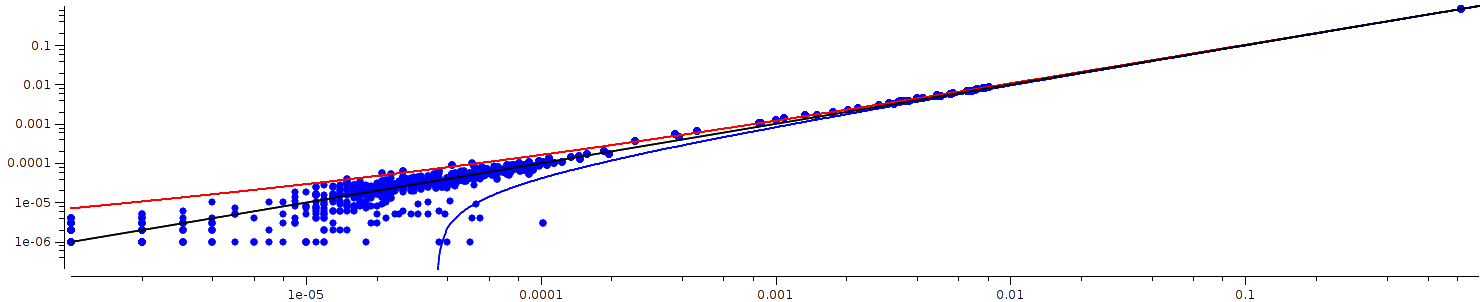}
 \label{sr2_1}
 }
 \centering
 \subfigure[Patterns of depth 2]{
\includegraphics[height=4cm,width=0.45\textwidth]{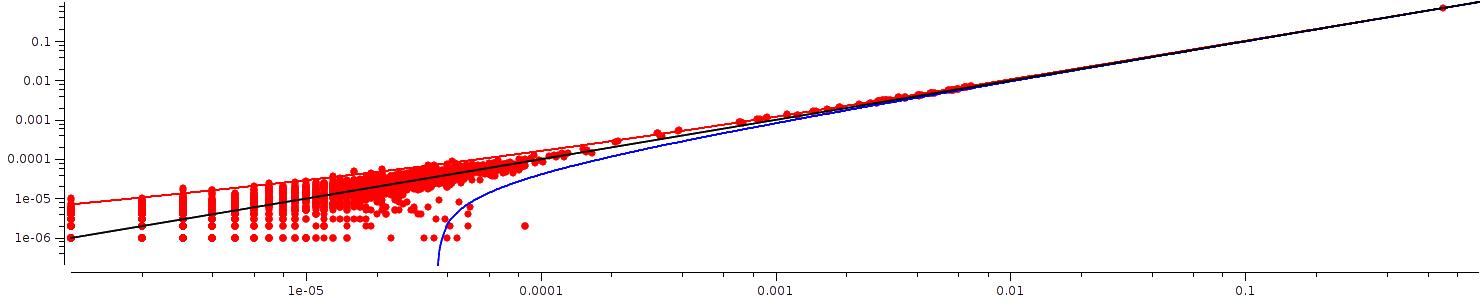}
 \label{sr2_2}
 }
  \centering
 \subfigure[Patterns of depth 3]{
\includegraphics[height=4cm,width=0.45\textwidth]{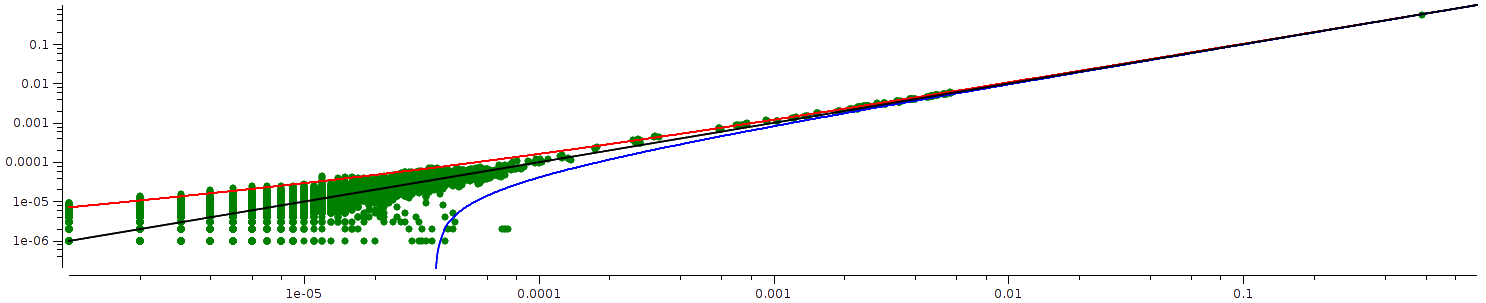}
 \label{sr2_3}
 }
\caption[Result on synthetic data with an Ising model.]{Data were generated with an Ising distribution. After fitting with an Ising model, we show the comparison between observed and predicted probabilities of monomials in (a). (b), (c), (d) presents the comparison of predicted and observed probabilities of patterns of depth 1,2 and 3 respectively. In the 4 plots: the x-axis represent the observed probabilities and the y-axis represent the predicted probabilities. The estimated Kullback-Leibler divergence is 0.0107.}
 \label{Sisingsynthetic}
\end{figure}
 \item Spatio-temporal, 40 neurons, $R = 2$ ($NR = 80$): Pairwise model with delays (\ref{HPR}). 
 Fig. \ref{Spairwisesynthetic}
 
  \begin{figure}[!]
 \centering
   \centering
    \subfigure[Monomials averages]{
\includegraphics[height=4cm,width=0.45\textwidth]{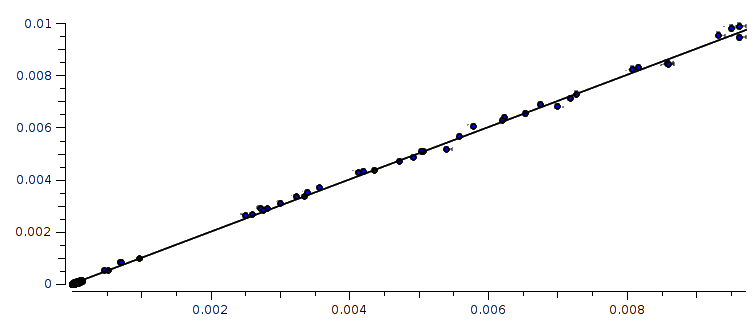}
 \label{si_monomials}
 }
 \subfigure[Patterns of depth 1]{
\includegraphics[height=4cm,width=0.45\textwidth]{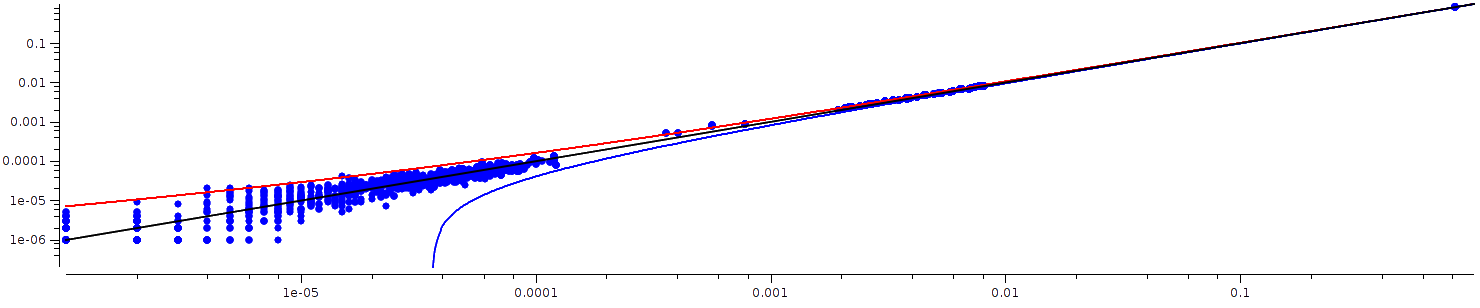}
 \label{si_1}
 }
 \centering
 \subfigure[Patterns of depth 2]{
\includegraphics[height=4cm,width=0.45\textwidth]{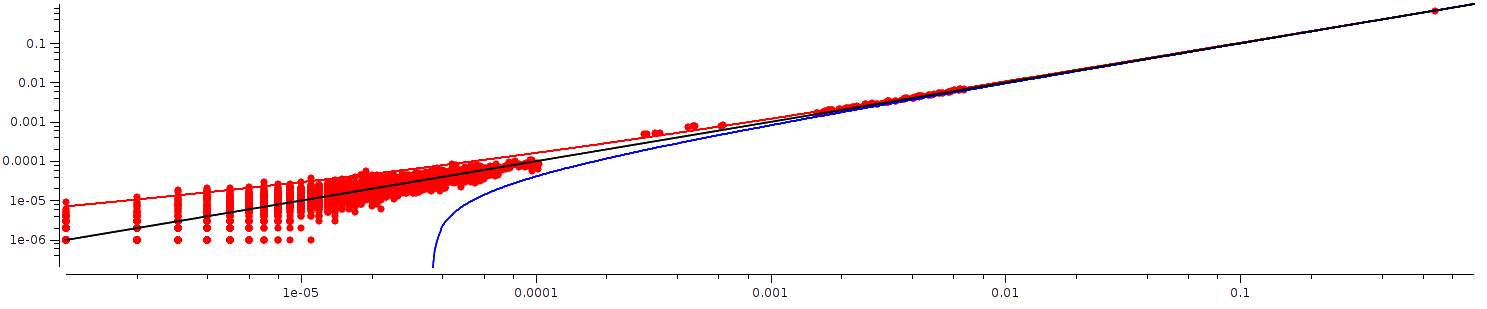}
 \label{si_2}
 }
  \centering
 \subfigure[Patterns of depth 3]{
\includegraphics[height=4cm,width=0.45\textwidth]{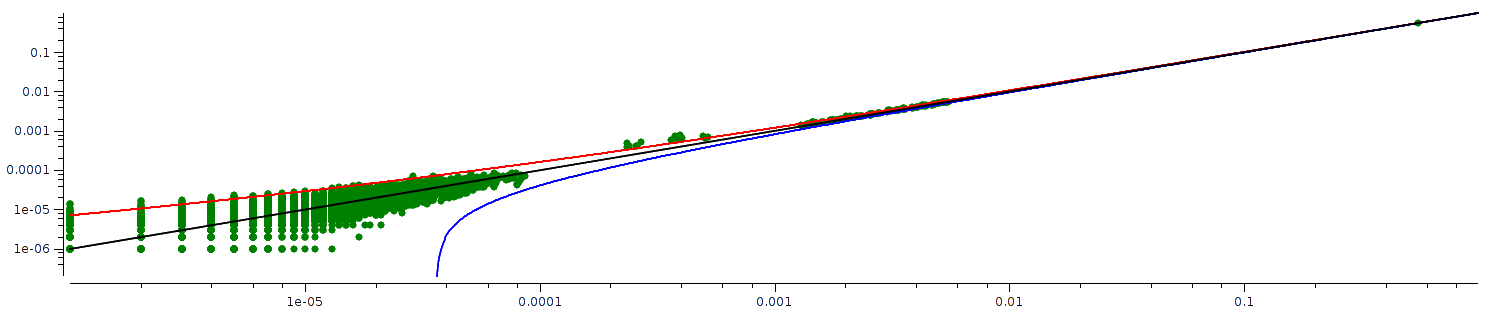}
 \label{si_3}
 }
\caption[Result on synthetic data with a pairwise model.]{Data were generated with a pairwise distribution of range $R=2$. After fitting with a pairwise model of Range $R=2$, we show the comparison between observed and predicted probabilities of monomials in (a). (b), (c), (d) presents the comparison of predicted and observed probabilities of patterns of depth 1,2 and 3 respectively. In the 4 plots: the x-axis represent the observed probabilities and the y-axis represent the predicted probabilities. The estimated Kullback-Leibler divergence is 0.0174.}
 \label{Spairwisesynthetic}
\end{figure}

\end{enumerate}


\subsection{The performance on real data}
Here we show the inferring of MaxEnt distribution on real spike trains. We analyzed a set of 20 and 40 neurons\footnote{40 is the maximal number of neurons in this data set} (courtesy of M. J. Berry and O. Marre) with spatial and spatio temporal constraints. Data are binned at 20 ms. We show the confidence plots and an example of convergence curves using the Hellinger Distance. The goal here is to check the goodness of fit not only for spatial patterns (as done in \cite{schneidman-berry-etal:06,pillow-shlens-etal:08,ganmor-segev-etal:11b,ganmor-segev-etal:11a}), but also for spatio-temporal patterns.

\begin{figure}[!]
\centering
\subfigure[Convergence during the parallel update]{
 \centering
 \includegraphics[width=1\textwidth, height=0.3\textheight]{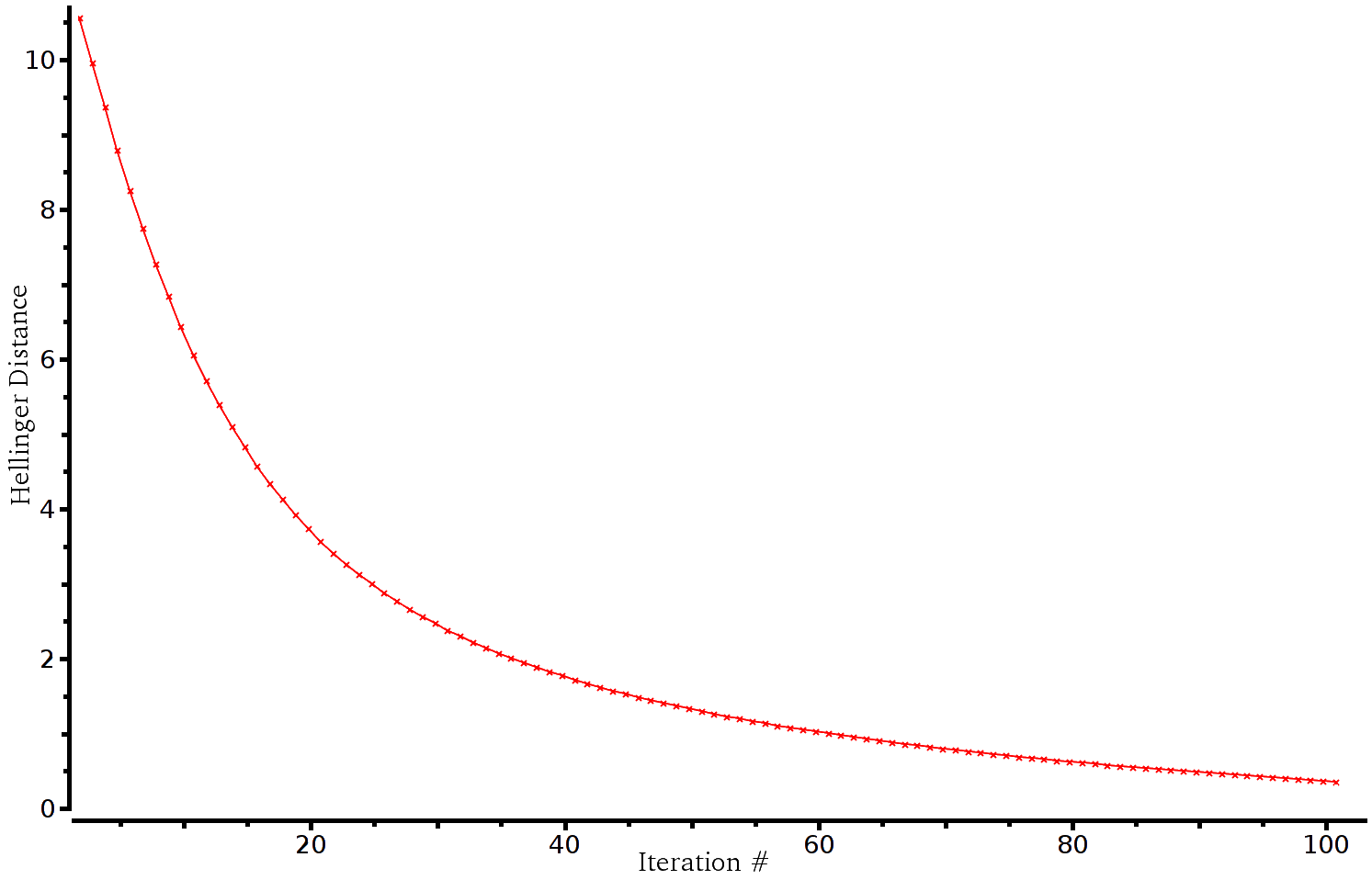}
 \label{cpar}
}
\subfigure[Convergence during the parallel update]{
 \centering
 \includegraphics[width=1\textwidth, height=0.3\textheight]{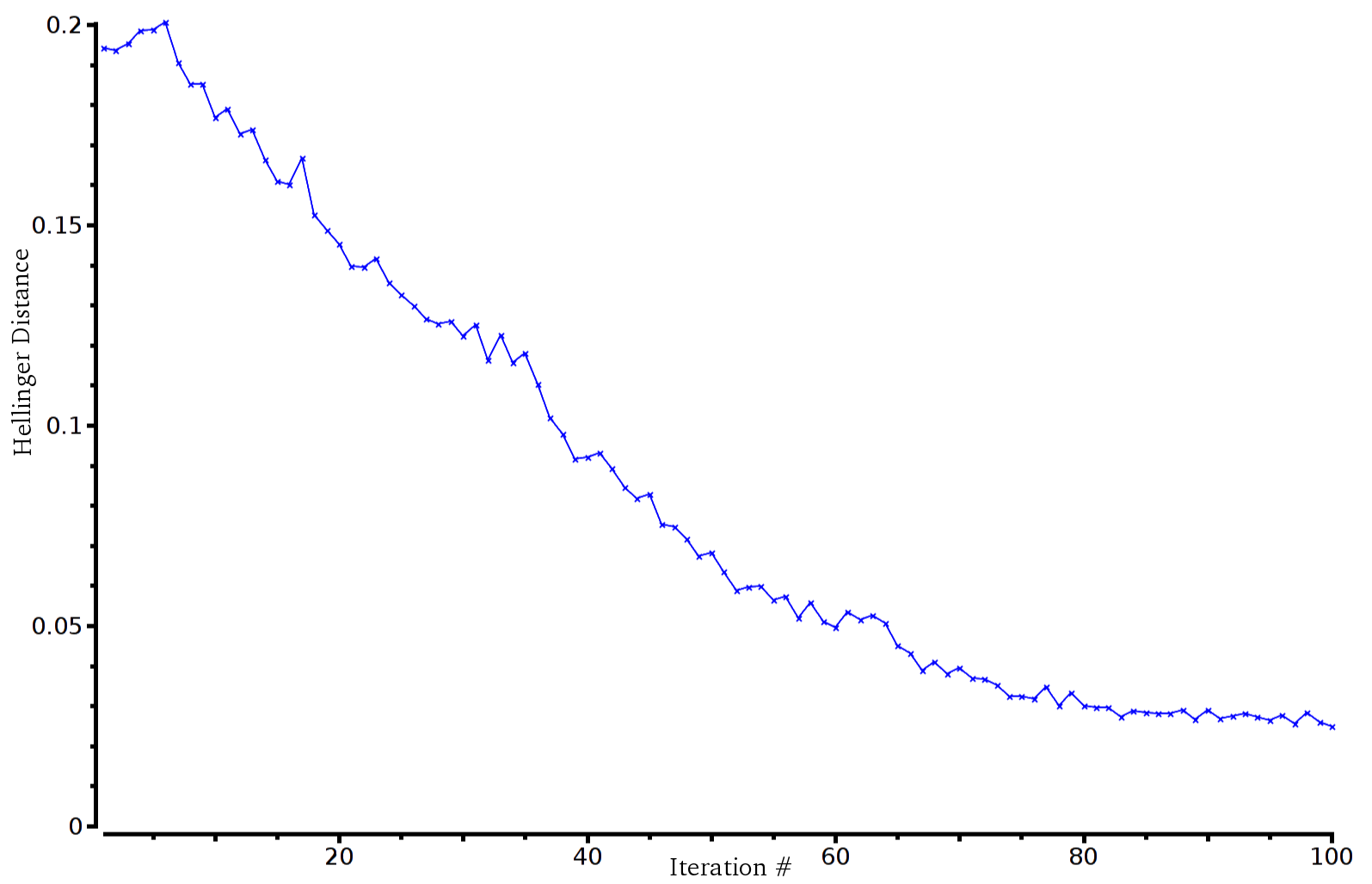}
 \label{cseq}
}
\caption[Convergence during updating the parameters]{Evolution of the Hellinger distance during the parallel (a) and the sequential (b) update in the case of modeling a real data set with a pairwise model of range $R=2$. The parallel update provides a fast convergence however it is steady after a hundred of iterations. Then we iterate the sequential algorithm.}
\label{convergence}
\end{figure}

Figure \ref{convergence} show the evolution of the Hellinger distance during parameters update both in parallel and sequential update process.

\begin{figure}[!]
 \centering
   \centering
 \subfigure[Monomials]{
\includegraphics[height=4cm,width=0.45\textwidth]{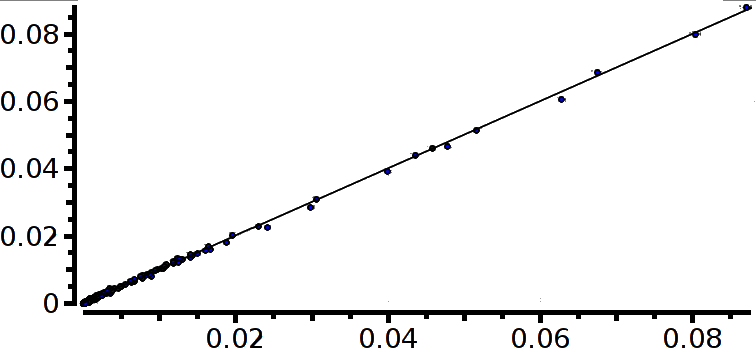}
 \label{monoms_ising_20}
 }
\centering
\subfigure[Patterns of range 1]{
 \centering
 \includegraphics[height=4cm,width=0.45\textwidth]{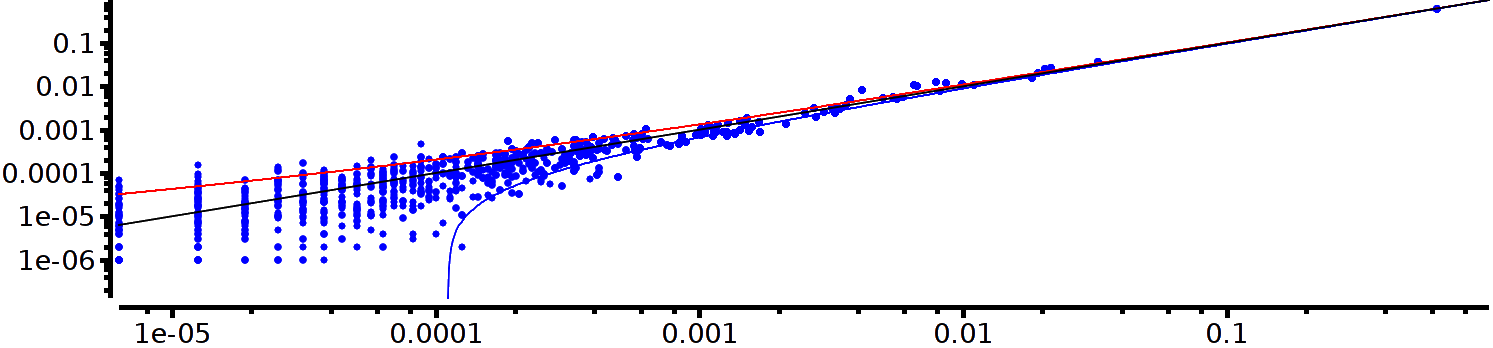}
 \label{i20_1}
}
\subfigure[Patterns of range 2]{
 \centering
 \includegraphics[height=4cm,width=0.45\textwidth]{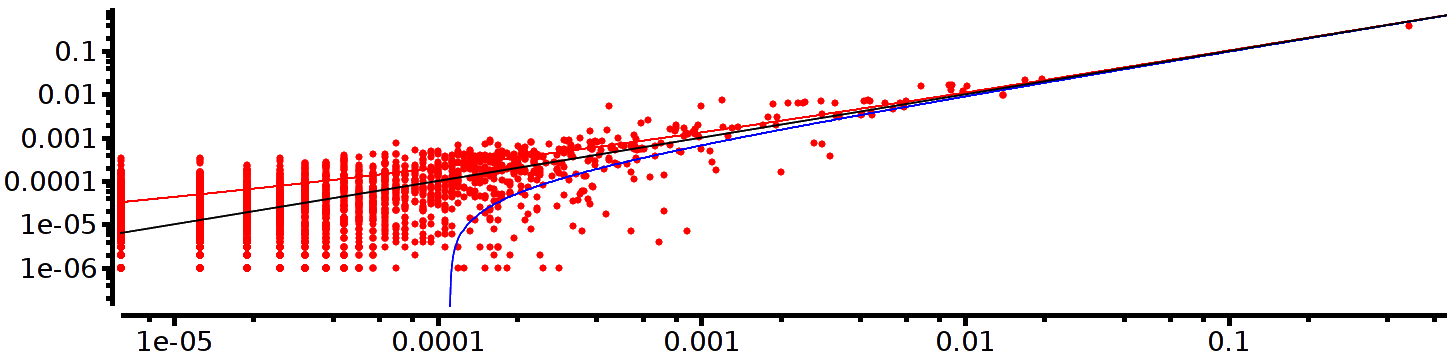}
 \label{i20_2}
}
\subfigure[Patterns of range 3]{
 \centering
 \includegraphics[height=4cm,width=0.45\textwidth]{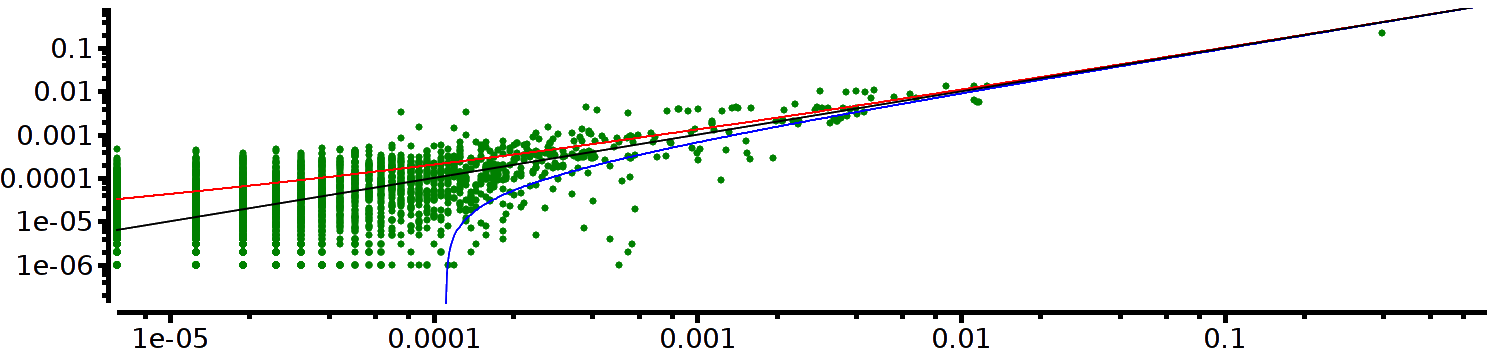}
 \label{i20_3}
}
\caption[Result on real data with an Ising model.]{A 20 neurons data set binned at 20 ms with an Ising model. After fitting, we show the comparison between observed (in the real spike train) and predicted average values of monomials in (a). (b), (c) and (d) present the comparison of predicted and observed probabilities for patterns of range 1,2 and 3 respectively. In the (a), (b), (c) and (d): the x-axis represents the observed probabilities and the y-axis represents the predicted probabilities. The computation time is equal to 18 hours on a small cluster of 64 processors (around 5 min per iteration). The estimated Kullback-Leibler divergence is 0.307.}
 \label{ising_20_elad}
\end{figure}

\begin{figure}[!]
 \centering
   \centering
 \subfigure[Monomials averages]{
\includegraphics[height=4cm,width=0.45\textwidth]{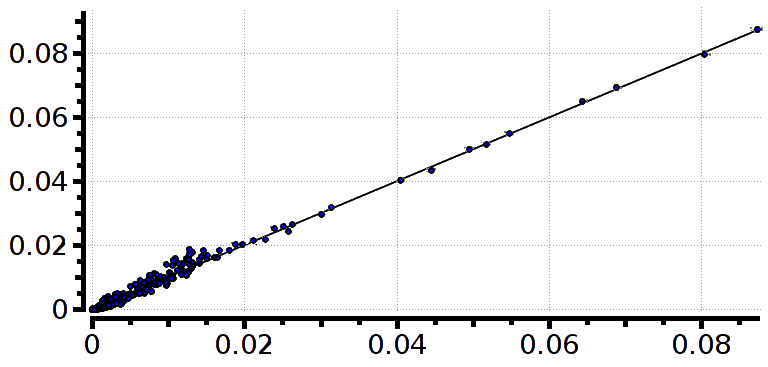}
 \label{monoms_ising_20}
 }
\centering
\subfigure[Patterns of range 1]{
 \centering
 \includegraphics[height=4cm,width=0.45\textwidth]{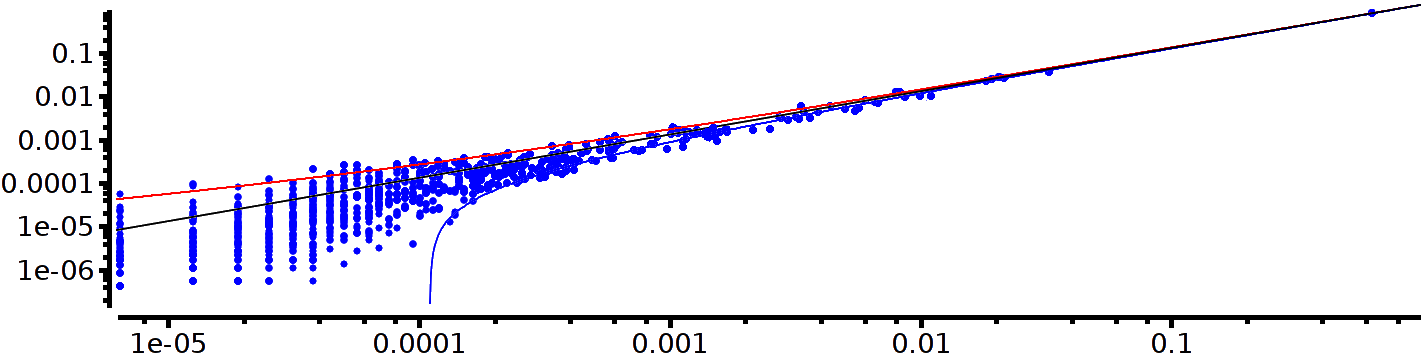}
 \label{i20_1}
}
\subfigure[Patterns of range 2]{
 \centering
 \includegraphics[height=4cm,width=0.45\textwidth]{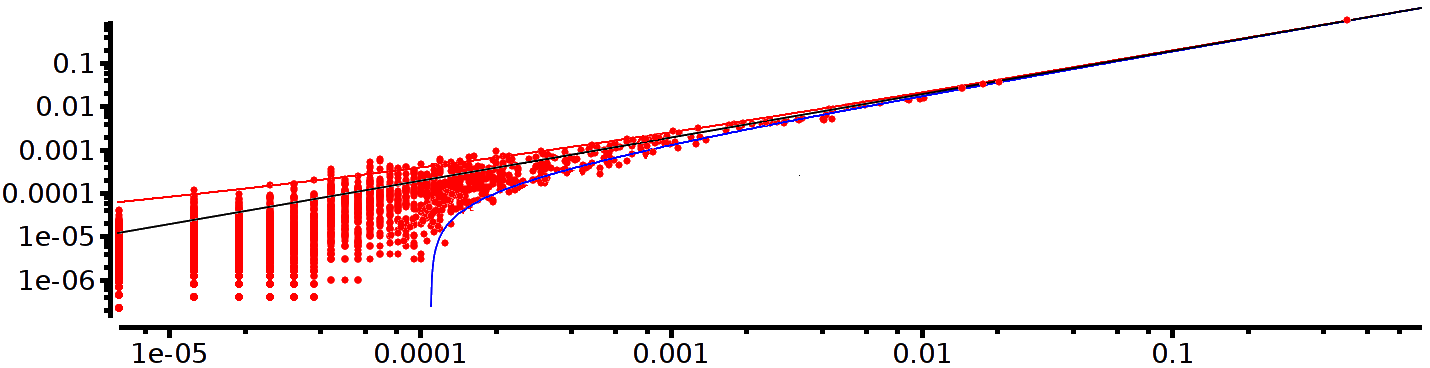}
 \label{i20_2}
}
\subfigure[Patterns of range 3]{
 \centering
 \includegraphics[height=4cm,width=0.45\textwidth]{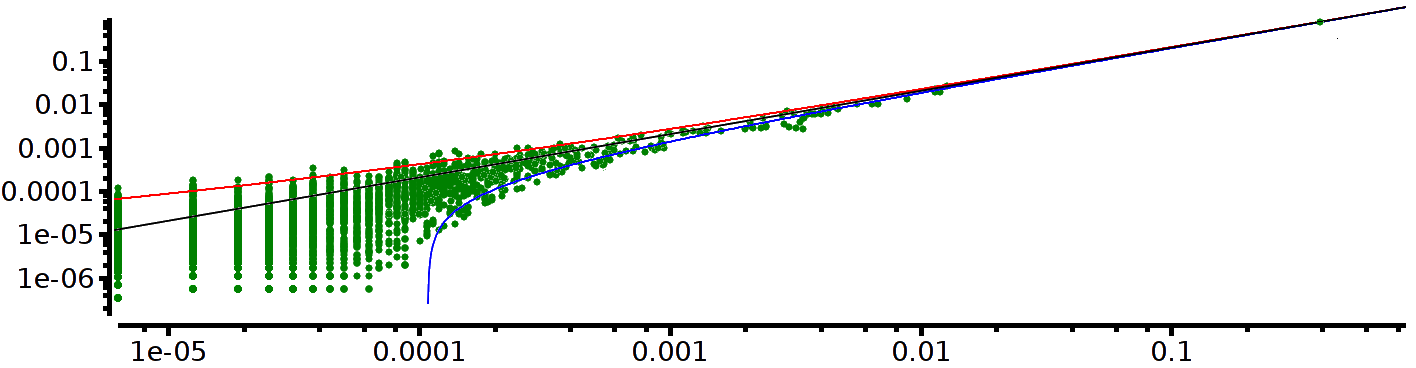}
 \label{i20_3}
}
\caption[Result on real data with a pairwise model.]{A 20 neurons data set binned at 20 ms with a pairwise model of range 2. After fitting, we show the comparison between observed (in the real spike train) and predicted average values of monomials in (a). (b), (c) and (d) present the comparison of predicted and observed probabilities for patterns of range 1,2 and 3 respectively. In the (a), (b), (c) and (d): the x-axis represents the observed probabilities and the y-axis represents the predicted probabilities. The computation time is equal to 40 hours on a small cluster of 64 processors (around 12 min per iteration).The estimated Kullback-Leibler divergence is 0.281.}
 \label{r2_20_elad}
\end{figure}

After estimating the parameters of an Ising and pairwise model of range $R=2$ on a set of 20 neurons, we evaluate the confidence plots. Figures \ref{ising_20_elad} and \ref{r2_20_elad} show respectively the confidence plots for patterns of range 1,2 and 3 after fitting with an Ising model and Pairwise model of range $R=2$. Our results on 20 neurons confirm the observations made in \cite{vasquez-marre-etal:12} for $N=5, R =2$ : a pairwise model with memory performs quite better than an Ising model to explain spatio-temporal patterns.

\begin{figure}[!]
 \centering
   \centering
 \subfigure[Monomials averages]{
\includegraphics[height=4cm,width=0.45\textwidth]{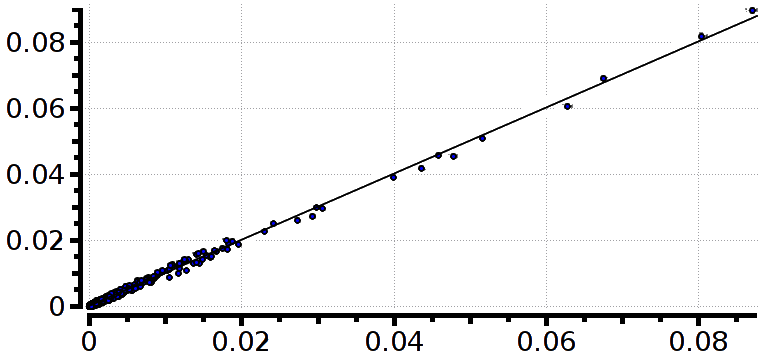}
 \label{monoms_ising_40}
 }
\centering
\subfigure[Patterns of range 1]{
 \centering
 \includegraphics[height=4cm,width=0.45\textwidth]{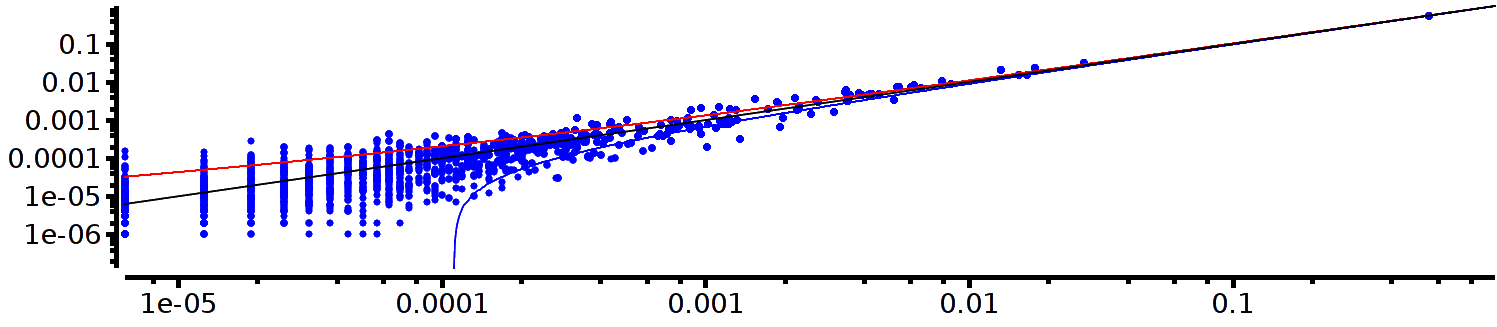}
 \label{i40_1}
}
\subfigure[Patterns of range 2]{
 \centering
 \includegraphics[height=4cm,width=0.45\textwidth]{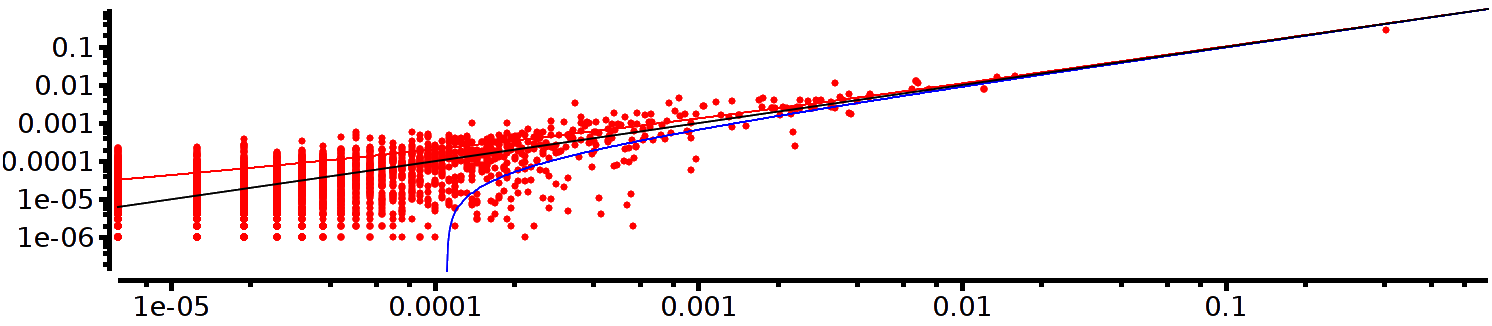}
 \label{i40_2}
}
\subfigure[Patterns of range 3]{
 \centering
 \includegraphics[height=4cm,width=0.45\textwidth]{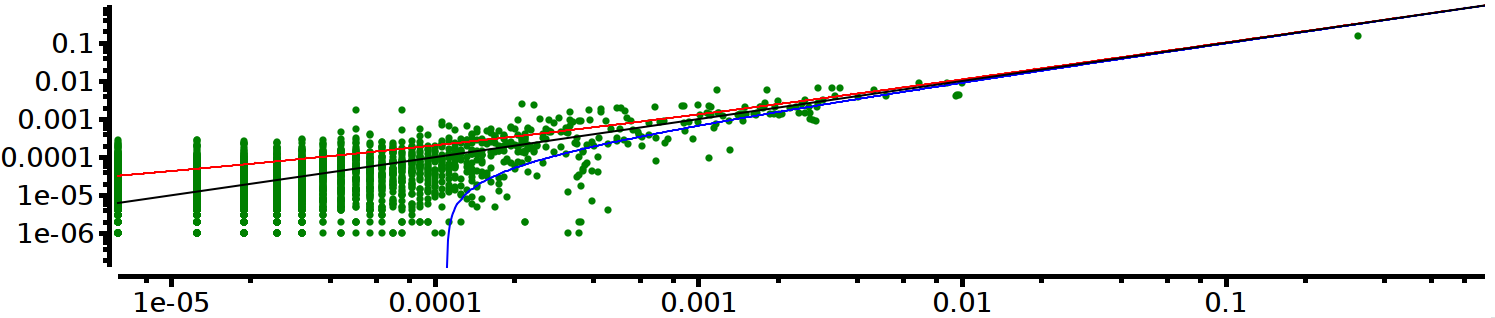}
 \label{i40_3}
}
\caption[Result on synthetic data with an Ising model.]{A 40 neurons data set binned at 20 ms with an Ising model. After fitting, we show the comparison between observed (in the real spike train) and predicted average values of monomials in (a). (b), (c) and (d) present the comparison of predicted and observed probabilities for patterns of range 1,2 and 3 respectively. In the (a), (b), (c) and (d): the x-axis represents the observed probabilities and the y-axis represents the predicted probabilities. The computation time is equal to 3 days on a small cluster of 64 processors (around 21 min per iteration). The estimated Kullback-Leibler divergence is 0.930.}
 \label{ising_40_elad}
\end{figure}

\begin{figure}[!]
 \centering
   \centering
 \subfigure[Monomials averages]{
\includegraphics[height=4cm,width=0.45\textwidth]{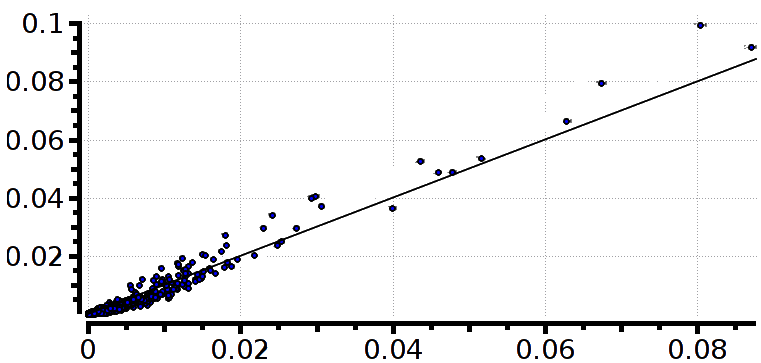}
 \label{monoms_r2_40}
 }
\centering
\subfigure[Patterns of range 1]{
 \centering
 \includegraphics[height=4cm,width=0.45\textwidth]{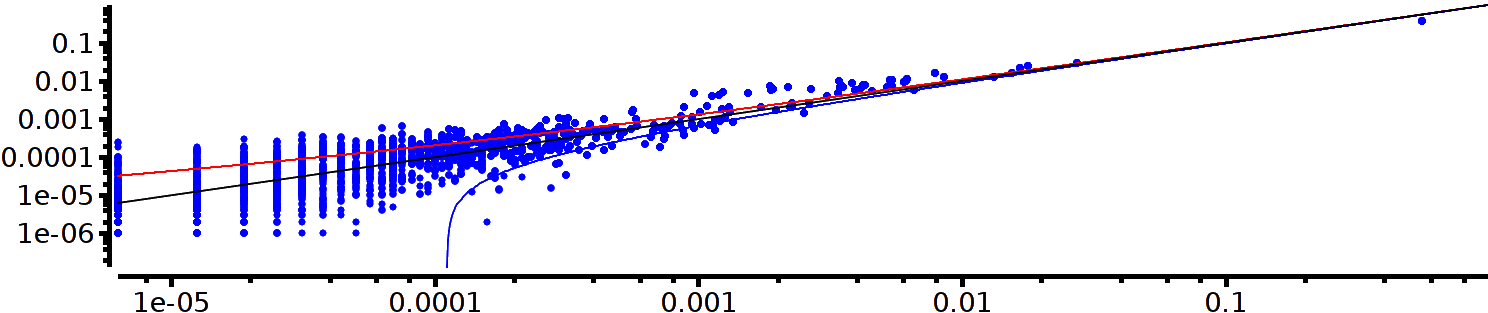}
 \label{r2_40_1}
}
\subfigure[Patterns of range 2]{
 \centering
 \includegraphics[height=4cm,width=0.45\textwidth]{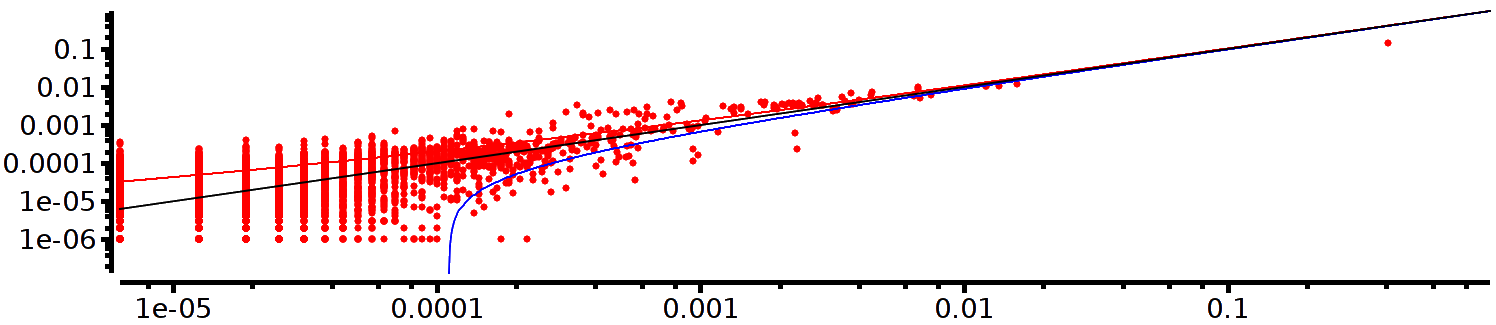}
 \label{r2_40_2}
}
\subfigure[Patterns of range 3]{
 \centering
 \includegraphics[height=4cm,width=0.45\textwidth]{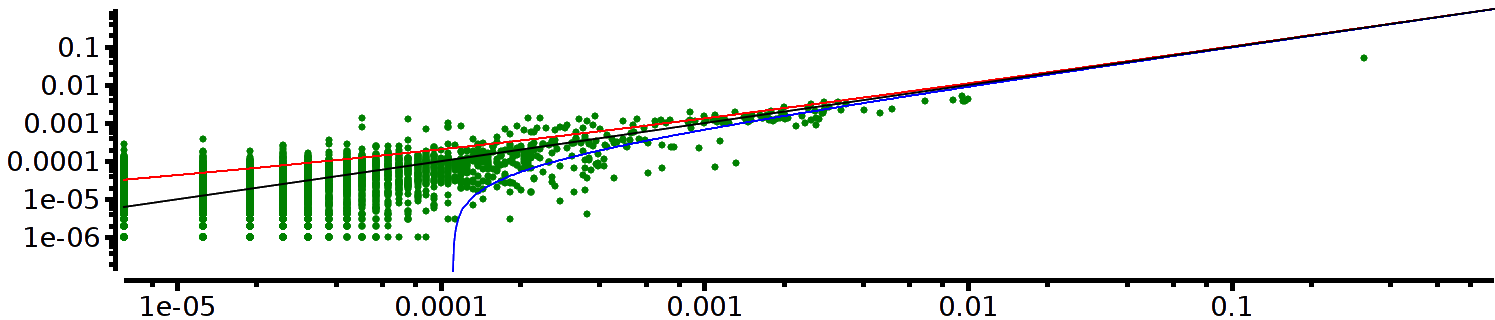}
 \label{r2_40_3}
}
\caption[Result on real data with a pairwise model.]{A 40 neurons data set binned at 20 ms with a pairwise model of range 2. After fitting, we show the comparison between observed (in the real spike train) and predicted average values of monomials in (a). (b), (c) and (d) present the comparison of predicted and observed probabilities for patterns of range 1,2 and 3 respectively. In the (a), (b), (c) and (d): the x-axis represents the observed probabilities and the y-axis represents the predicted probabilities. The computation time is equal to 7 days on a small cluster of 64 processors (around 47 min per iteration). The estimated Kullback-Leibler divergence is 0.983.}
 \label{r2_40}
\end{figure}

We then made the same analysis for 40 neuron. Figures \ref{ising_40_elad} and \ref{r2_40} show respectively the confidence plots for patterns of range 1,2 and 3 after fitting with an Ising model and Pairwise model of range $R=2$. In this case, we were not able to obtain a good convergence for $N=40, R=2$. This is presumably due to the insufficient length of the data set which does not allow us to estimate accurately the probability of some monomials. This aspect is discussed in the next section. 

\section{Discussion and conclusion}

The method shows better performances for synthetic data than for real data although we did not make extensive studies for real data.
The main reason, we believe, is that in the second case we don't know the form of the potential. As a consequence, we stick at existing canonical forms of potentials e.g. Ising and pairwise. The main problem with this approach is that the number of parameters to estimate dramatically growths with $NR$. The increase is moderate for the Ising model ($N$ rates + $\frac{N(N-1)}{2}$
symmetric pairwise couplings) but it becomes prohibitively large even for pairwise range $R$ models. On the opposite, our analysis of synthetic data used a relatively small number of parameters to fit. 
 
The large number of parameters has 2 drawbacks: the increasing of computation time and errors in the estimation. Let us comment on the second problem. It is not intrinsic to our method; it is neither intrinsic to MaxEnt; this is a well known problem which arises already when doing linear regression analysis. Increasing the number of parameters may eventually lead to catastrophic estimations where the addition of degree of freedom can seriously hinder the resolution.

In the case of MaxEnt the situation can be described as follows. 
We generate a finite raster $\omega_0^T$ from a known distribution $\mbs$ with a potential of the form (\ref{Hh}). Denote $\mbs\bra{\m}$ the vector with entries $\mbs\bra{m_l}$ and $\pTo{\m}$ the vector with entries $\pTo{m_l}$. From (\ref{dPxl}) we have $\mbs\bra{\m} = \nabla_{\blambda^\ast} \cP$. This exact solution is obtained when the Gibbs distribution $\mbs$ can be exactly sampled, namely, for an infinite raster. For a finite raster, if  $T$ is large enough to apply the central limit theorem, the empirical distribution $\pTo{\m}$ is Gaussian with mean $\mb\bra{\m}$ and covariance $\frac{1}{T} \chi$ given by (\ref{FDT}). We have therefore $\pTo{\m} = \mbs\bra{\m} + \bbeta$ where $\bbeta$ is centered Gaussian with covariance $\frac{1}{T} \chi$. Solving (\ref{dPxl}) where the exact probability $\mbs$ is replaced by the empirical one $\pi_{\omega}^{(T)}$, one obtains an approximate solution of $\blambda$, $\blambda^\ast$ with :
$\blambda=\blambda^\ast + \beps$,
where
$
\nabla_{\blambda} \cP= \pTo{\m}.
$
Therefore, $\nabla_{\blambda} \cP= \mbs\bra{\m} + \bbeta = \nabla_{\blambda^\ast + \beps}\cP $
$= \nabla_{\blambda^\ast} \cP + \beps \chi + O(\|\beps\|^2)$. Hence, $\beps = \chi^{-1} \bbeta$. $\chi$ is invertible since $\cP$ is convex.

The fluctuations
of the estimated solution $\blambda$ around the exact solution $\blambda^\ast$ are therefore Gaussian, centered, with covariance $
 \Exp{\epsilon.\tilde{\epsilon}} = \Exp{\chi^{-1}.\bbeta.\tilde{\bbeta} .\tilde{\chi}^{-1}}
$.
 Since $\chi$ is symmetric we have $
 \Exp{\epsilon.\tilde{\epsilon}}= \chi^{-1}.\Exp{\bbeta.\tilde{\bbeta}}.\chi^{-1}=\frac{1}{T} \, \chi^{-1}$. 
We arrive therefore at the conclusion that the fluctuations on the estimated coefficients $\blambda$
are highly constrained by the convexity of the pressure, as expected. Mathematically, everything goes nicely since $\cal P$ is convex. However, it may happen that $\cal P$ is quite flat in some directions/monomials. Then small errors will be largely amplified. Therefore, when considering potentials of the form (\ref{Hh}) it is expected that some terms (monomials) not only are irrelevant, but also dramatically deteriorate the estimation problem, introducing almost zero eigenvalues in $\chi$. This is presumably what happened in Figure \ref{r2_40} where we were not able to obtain a good convergence for monomials averages.

At this stage, the main question is therefore: Can we have an idea of the potential shape from data before fitting the parameters? This question is not only related to the goodness of fit but, it is also a question of concept. Is is useful to represent a pairwise distribution for 40 neurons with nearly 2000 parameters? The idea would then be to filter irrelevant monomials.
 For that a feature selection method is useful and should complement this work. There are many directions we can take in the favor of the features selection. For instance, selecting the features on threshold (\cite{rosenfeld-carbonell-etal:94,Koeling:00}), using a $\chi^2$ method (\cite{chen-rosenfeld:99})  as well as incremental feature selection algorithm (\cite{berger-etlal:96}, \cite{zou-lide:03}). Other methods based on periodic orbit sampling (\cite{cofre-cessac:13b}) and information geometry (\cite{hiroyuki-amari:01a, amari:01}) are under current investigation. 


We have presented a method to fit the parameters of MaxEnt distribution with spatio-temporal constraints. In the process of exploring the dynamics of neural data, we hypothesize the model, fit it and finally judge the quality of the suggested model. Hence, this work is positioned as an important intermediate step in the neural coding using the MaxEnt framework, opening the door for analyzing the dynamics of large networks being not limited to spatial and/or traditional MaxEnt models. 

Finally, we would like to highlight two points that should be investigated in further studies:

\begin{itemize}
 \item The effect of binning. In many experimental studies data is binned.
Basically, binning was used in order to account for time spiking sensitivity, which is not the same for all the biological neural networks.
 For instance, \cite{schneidman-berry-etal:06} used 20 ms of binning for retinal spike trains. In the present paper, we have used the same as these authors but we have not considered
the effect of binning on our statistical estimations. This is certainly a matter of further investigations, especially because,
to our best knowledge no systematic study on binning effects on statistics has been done. In particular, three distinct
dimensions should be considered:
\begin{itemize}
 \item The statistical dimension: How does binning biases statistics ? Could binning introduce spurious effects such as e.g. creating fallacious long range correlations?
 \item The computational dimension: how does the performance of the algorithm change with the bin size?
 \item The biological dimension: cross-correlograms are not the same in all brain areas. So optimal bin size is expected to depend on the investigated area.
\end{itemize}
 \item Maximum Entropy: There are several methods now in use to model the spatio-temporal correlations in ensembles of neurons. The generalized linear model (GLM) approach uses maximum likelihood and point-process to assess connectivity (e.g., \cite{pillow-shlens-etal:08}). Reverse correlation methods can also work well (e.g., \cite{chichilnisky:01}). Finally, there are causality metrics like Granger causality or transfer entropy (\cite{zhaohui-xiaoli:13}). Some of these methods have been compared in \cite{truccolo-etal:09}, but further investigations should be helpful, starting from synthetic data where statistics is under good control.  Especially, how does Maximum entropy perform compared to these others methods?
\end{itemize}

Our method allow to investigate these two questions on numerical grounds although such an investigation should be completed by mathematical insights, using the properties of spatio-temporal Gibbs distributions.

\newpage
\section{List of symbols}

\begin{tabular}{ll}
 $\omega_i(n)$ & Spike event\\
 $\omega(n)$ & Spike pattern \\
 $\omega_{n_1}^{n_2}$ & Spike block \\
 $\omega$ & Spike train\\
 $T$ & Length (in time) of the spike train \\
 $N$ & Number of neurons \\
 $R$ & Model range\\
 $D$ & Model memory ($R=D-1$) \\
 $m_l(\omega)$ & Monomial number $l$ \\
 $\m$ & Vector of monomials \\
 $L$ & Total number of parameters (monomials) in the model \\
 $\lambda_l$ & Parameter number $l$ \\
 $\blambda$ & Parameters vector \\
 ${\cal H}$ & Gibbs potential \\
 $\Zb$ & Partition function \\
 ${\cal S}$  & Entropy \\
 ${\cal P}$ & Topological pressure \\
 $\pi_{\omega}^{(T)}$ & Empirical probability measured on the spike train $\omega$ of length $T$\\
 $\mu_{\blambda}$ & Gibbs density with parameters $\blambda$ \\
 ${\cal M}$ & Set of invariant probabilities \\
 $\delta_l = \lambda_l'-\lambda_l$ & Learning rate or the value by which we update the parameters $\lambda_l$ \\
 $\bdelta$ & Vector of learning rates \\
 $d_{KL}$ & Kullback-Leibler divergence \\
 $C_{jk}$ & Correlation between two monomials $j$ and $k$ \\
 $\chi$ & Hessian matrix (second derivative of the pressure) \\
 $\Delta$ & Root sum square of the learning rates \\
 $\bbeta$ & Fluctuations on the monomials averages \\
 $\beps$ & Fluctuations on the parameters (relaxation) \\
\end{tabular}

\acknowledgements{Acknowledgments}
We thank the reviewers for helpful remarks and constructive criticism. We also warmly acknowledge M. J. Berry, O. Marre for providing us MEA recordings from the retina and G. Tkacik who provided us the references \cite{dudik-phillips-etal:04,broderick-dudik-etal:07} and helped us in the algorithm design. This work was partially supported by the  ERC-NERVI number 227747, KEOPS ANR-CONICYT, and European FP7 projects RENVISION (FP7-600847), BRAINSCALES (FP7-269921).

\conflictofinterests{Conflicts of Interest}

The authors declare no conflicts of interest.

%

\end{document}